\documentclass[twocolumn]{aastex6}
\usepackage{amsmath}
\usepackage{CJK}

\makeatletter
\newcommand{\bdv}[1]{\mbox{\boldmath$#1$}}
 \newcommand{\Rmnum}[1]{\expandafter\@slowromancap\romannumeral #1@}
\def\p{{\rm p}}
\def\e{{\rm E}}
\def\l{{\rm L}}
\def\s{{\rm S}}
\def\rel{{\rm rel}}
\def\au{{\rm AU}}
\def\en{{\rm E,N}}
\def\ee{{\rm E,E}}
\def\gc{{\rm gc}}
\def\mas{{\rm mas}}
\def\sat{{\rm sat}}
\def\hel{{\rm hel}}
\def\geo{{\rm geo}}
\def\kpc{{\rm kpc}}
\def\dif{{\rm d}}
\def\min{{\rm min}}
\def\max{{\rm max}}
\def\microm{\mu{\rm m}}
\def\kms{{\rm km~s^{-1}}}
\def\masyr{{\rm mas~yr^{-1}}}
\def\sub{{\rm sub}}
\def\jup{{\rm J}}
\def\b2d{{\rm b2d}}
\def\Dparm{D_{8.3}}
\def\Rgc{R_{\rm GC}}
\def\data{{\rm Data}}

\shorttitle{Galactic Distribution of Planets. \Rmnum{1}. Methods \& First Sample}
\shortauthors{Zhu et al.}

\begin{document}
\begin{CJK*}{UTF8}{gbsn}

\title{Toward A Galactic Distribution of Planets. \Rmnum{1}.\\Methodology \& Planet Sensitivities of the 2015 High-Cadence \emph{Spitzer} Microlens Sample}

\author{
Wei~Zhu~(祝伟)\altaffilmark{1,16,17},
A.~Udalski\altaffilmark{2,18},
S.~Calchi~Novati\altaffilmark{3,4,16},
S.-J.~Chung\altaffilmark{5,6,17},
Y.~K.~Jung\altaffilmark{7,17},
Y.-H.~Ryu\altaffilmark{5,17},
I.-G.~Shin\altaffilmark{7,17},
A.~Gould\altaffilmark{1,5,8,16,17},
C.-U.~Lee\altaffilmark{5,6,17},
M.~D.~Albrow\altaffilmark{9,17},
J.~C.~Yee\altaffilmark{7,16,17} \\
AND \\
C.~Han\altaffilmark{10},
K.-H.~Hwang\altaffilmark{5},
S.-M.~Cha\altaffilmark{5,11},
D.-J.~Kim\altaffilmark{5},
H.-W.~Kim\altaffilmark{5},
S.-L.~Kim\altaffilmark{5,6},
Y.-H.~Kim\altaffilmark{5},
Y.~Lee\altaffilmark{5,11},
B.-G.~Park\altaffilmark{5,6},
R.~W.~Pogge\altaffilmark{1} \\
(KMTNet Collaboration) \\
R.~Poleski\altaffilmark{1,2},
J.~Skowron\altaffilmark{2},
P.~Mr{\'o}z\altaffilmark{2},
M.~K.~Szyma{\'n}ski\altaffilmark{2},
I.~Soszy{\'n}ski\altaffilmark{2},
P.~Pietrukowicz\altaffilmark{2},
S.~Koz{\l}owski\altaffilmark{2},
K.~Ulaczyk\altaffilmark{2,12},
M.~Pawlak\altaffilmark{2} \\
(OGLE Collaboration) \\
\and
C.~Beichman\altaffilmark{13},
G.~Bryden\altaffilmark{14},
S.~Carey\altaffilmark{15},
M.~Fausnaugh\altaffilmark{1},
B.~S.~Gaudi\altaffilmark{1},
C.~B.~Henderson\altaffilmark{14,19},
Y.~Shvartzvald\altaffilmark{14,19},
B.~Wibking\altaffilmark{1} \\
(Spitzer Team)
}

\email{zhu.908@osu.edu}

\altaffiltext{1}{Department of Astronomy, Ohio State University, 140 W. 18th Ave., Columbus, OH  43210, USA}
\altaffiltext{2}{Warsaw University Observatory, AI. Ujazdowskie 4, 00-478 Warszawa, Poland}
\altaffiltext{3}{IPAC, Mail Code 100-22, Caltech, 1200 E. California Blvd, Pasadena, CA 91125, USA}
\altaffiltext{4}{Dipartimento di Fisica ``E. R. Caianiello'', Universit\`a di Salerno, Via Giovanni Paolo II, 84084 Fisciano (SA), Italy}
\altaffiltext{5}{Korea Astronomy and Space Science Institute, 776 Daedeokdae-ro, Yuseong-Gu, Daejeon 34055, Korea}
\altaffiltext{6}{Korea University of Science and Technology, 217 Gajeong-ro, Yuseong-gu, Daejeon 34113, Korea}
\altaffiltext{7}{Harvard-Smithsonian Center for Astrophysics, 60 Garden St., Cambridge, MA 02138, USA}
\altaffiltext{8}{Max-Planck-Institute for Astronomy, K\"onigstuhl 17, 69117 Heidelberg, Germany}
\altaffiltext{9}{Department of Physics and Astronomy, University of Canterbury, Private Bag 4800 Christchurch, New Zealand}
\altaffiltext{10}{Department of Physics, Chungbuk National University, Cheongju 361-763, South Korea}
\altaffiltext{11}{School of Space Research, Kyung Hee University, Giheung-gu, Yongin, Gyeonggi-do, 17104, Korea}
\altaffiltext{12}{Department of Physics, University of Warwick, Gibbet Hill Road, Coventry, CV4~7AL, UK}
\altaffiltext{13}{NASA Exoplanet Science Institute, MS 100-22, California Institute of Technology, Pasadena, CA 91125, USA}
\altaffiltext{14}{Jet Propulsion Laboratory, California Institute of Technology, 4800 Oak Grove Drive, Pasadena, CA 91109, USA}
\altaffiltext{15}{Spitzer Science Center, MS 220-6, California Institute of Technology, Pasadena, CA 91125, USA}
\altaffiltext{16}{Spitzer Team}
\altaffiltext{17}{KMTNet Collaboration}
\altaffiltext{18}{OGLE Collaboration}
\altaffiltext{19}{NASA Postdoctoral Program Fellow}

\begin{abstract}
    We analyze an ensemble of microlensing events from the 2015 \emph{Spitzer} microlensing campaign, all of which were densely monitored by ground-based high-cadence survey teams. The simultaneous observations from \emph{Spitzer} and the ground yield measurements of the microlensing parallax vector $\bdv{\pi_\e}$, from which compact constraints on the microlens properties are derived, including $\lesssim$25\% uncertainties on the lens mass and distance.
With the current sample, we demonstrate that the majority of microlenses are indeed in the mass range of M dwarfs.
    The planet sensitivities of all 41 events in the sample are calculated, from which we provide constraints on the planet distribution function. 
    {In particular, assuming a planet distribution function that is uniform in $\log{q}$, where $q$ is the planet-to-star mass ratio, we find a $95\%$ upper limit on the fraction of stars that host typical microlensing planets of 49\%, which is consistent with previous studies.}
    Based on this planet-free sample, we develop the methodology to statistically study the Galactic distribution of planets using microlensing parallax measurements. 
    Under the assumption that the planet distributions are the same in the bulge as in the disk, we predict that $\sim$1/3 of all planet detections from the microlensing campaigns with \emph{Spitzer} should be in the bulge. This prediction will be tested with a much larger sample, and deviations from it can be used to constrain the abundance of planets in the bulge relative to the disk.
\end{abstract}


\section{Introduction} \label{sec:introduction}

The distribution of planets in different environments is of great interest. Studies have shown that the planet frequency may be correlated with the host star metallicity \citep[e.g.,][]{Santos:2001,Santos:2003,FischerValenti:2005,WangFischer:2015,Zhu:2016b}, the stellar mass \citep[e.g.,][]{Johnson:2010}, stellar multiplicity \citep[e.g.,][]{Eggenberger:2007,Wang:2014}, and exterior stellar environment \citep[e.g.,][]{Thompson:2013}. For this purpose, probing the planet distribution outside the Solar Neighborhood is important. In particular, the planet distribution in the Galactic bulge, given its unique environment, can provide an extra dimension to test and further develop our theories of planet formation.

Probing the distribution of planets in the Galactic bulge, or more generally, at all Galactic scales, is a unique application of Galactic microlensing, because of its independence on the flux from the planet host \citep{MaoPaczynski:1991,GouldLoeb:1992}. For example, \citet{Penny:2016} used an ensemble of 31 microlensing planets and found tentative evidence that the bulge might be deficient of planets compared to the disk. 

While microlensing is in principle sensitive to planets at various Galactic distances, the distance determination of any given microlensing event is nontrivial. This is because, in the majority of cases, the only relevant observable from the microlensing light curve is the Einstein timescale
\begin{equation}
    t_\e \equiv \frac{\theta_\e}{\mu_\rel}\ .
\end{equation}
Here $\mu_\rel$ is lens-source relative proper motion, and $\theta_\e$ is the angular Einstein radius,
\begin{equation}
\theta_\e \equiv \sqrt{\kappa M_\l \pi_\rel};\quad \kappa \equiv \frac{4G}{c^2\au} \simeq 8.14 \frac{\mas}{M_\odot}\ ,
\end{equation}
where $M_\l$ is the lens mass, $\pi_\rel \equiv \au(D_\l^{-1}-D_\s^{-1})$ is the lens-source relative parallax, and $D_\l$ and $D_\s$ are distances to the lens and the source (i.e., the star being lensed), respectively. In planetary events, $\theta_\e$ is usually also measurable through the so-called finite-source effect \citep{Yoo:2004}, in addition to two parameters that characterize the planet itself: the planet/star mass ratio $q$ and the planet/star separation $s$ in units of $\theta_\e$ \citep{GaudiGould:1997b}. There nevertheless remains a degeneracy between the lens mass and lens distance (assuming the source is in the bulge, which is almost always the case). The difficulty in precisely determining the lens distance is a significant weakness of ground-based microlensing in determining the Galactic distribution of planets, as has been demonstrated by \citet{Penny:2016}.

The most efficient way to determine or better constrain the lens distance $D_\l$ is by measuring the so-called microlens parallax vector $\bdv{\pi_\e}$
\begin{equation}
    \bdv{\pi_\e} \equiv \frac{\pi_\rel}{\theta_\e} \frac{\bdv{\mu_\rel}}{\mu_\rel}\ ,
\end{equation}
which can be effectively achieved by simultaneously observing the same event from at least two well-separated ($\mathcal{O}(1~\au)$) observatories \citep{Refsdal:1966,Gould:1994}. This is because, for typical Galactic microlensing events, the projected Einstein radius on the observer plane,
\begin{equation}
    \tilde{r}_\e = \frac{\au}{\pi_\e}\ ,
\end{equation}
is of order $\sim10~\au$, and thus observers separated by $\sim$1~AU would see considerably different light curves of the same microlensing event. 
For events with $\theta_\e$ measurements, including most planetary events, most binary events, and relatively rare single-lens events, the measurements of $\bdv{\pi_\e}$ directly yield the lens mass and lens-source relative parallax
\begin{equation} \label{eqn:mass}
    M_\l = \frac{\theta_\e}{\kappa \pi_\e};\quad \pi_\rel = \theta_\e \pi_\e\ ,
\end{equation}
the latter being a good proxy for distinguishing disk and bulge lenses (see Section~\ref{sec:methods}). For the great majority of single-lens events, $\theta_\e$ cannot be measured from the microlensing light curve, but the lens distribution ($M_\l$ and $\pi_\rel$) can be much more tightly constrained once $\bdv{\pi_\e}$ is measured, as first pointed out by \citet{Han:1995}.

For this reason, the \emph{Spitzer} Space Telescope has been employed for microlensing \citep{Dong:2007,prop2013,prop2014,prop2015a,prop2015b,prop2016}.
The 2014 \emph{Spitzer} microlensing experiment served as a pilot program that successfully demonstrated the ability to measure microlens parallax using \emph{Spitzer} \citep{Udalski:2015a,Yee:2015a,SCN:2015a,Zhu:2015a}. Starting in 2015, the main goal of \emph{Spitzer} microlensing campaigns became measuring the Galactic distribution of planets \citep{SCN:2015a,Yee:2015b}.

It is by no means trivial to organize \emph{Spitzer} and ground observations to enable a measurement of the Galactic distribution of planets that is unbiased by observational decisions. On the one hand, microlensing events must be chosen for \emph{Spitzer} observations very carefully in order to maximize both the sensitivity to planets of the whole sample and the probability that these observations will actually lead to a microlens parallax measurement. On the other hand, these observational decisions cannot in any way be influenced by whether planets have (or have not) been detected. The first objective requires that observational decisions make maximal use of available information, while the second means that a certain ``blindness'' to this information must be rigorously enforced. \citet{Yee:2015b} discussed in great detail how to optimize observations while enforcing this blindness, and a short summary is given in Section~\ref{sec:spitzer-obs}. Interested readers are urged to consult \citet{Yee:2015b} for more details.

Following the \citet{Yee:2015b} protocol, the 2015 \emph{Spitzer} microlensing campaign observed 170 microlensing events that were first found in the ground-based microlensing surveys, namely the Optical Gravitational Lensing Experiment \citep[OGLE,][]{Udalski:2003,Udalski:2015b} and the Microlensing Observations in Astrophysics \citep[MOA,][]{Bond:2001,Sako:2008}. In this work, we present analysis of 50 of them that fall within the footprints of OGLE and the prime fields of the newly established KMTNet \citep[Korean Microlensing Telescope Network,][]{Kim:2016}.

The present work is not aimed at directly answering how planets are distributed within the Galaxy. Instead, we develop a framework within which the above question can be ultimately addressed. It is nevertheless true that the 50 events in our sample, observed at $\sim$10 min cadence nearly continuously throughout year 2015, are more sensitive to planets than the majority of the remaining events in the 2015 \emph{Spitzer} sample. Another significant contributor to the overall planet sensitivity would be high-magnification events, which have nearly 100\% sensitivity to planets \citep{Griest:1998,Gould:2010} but are considerably rarer. These high-magnification events will be analyzed separately.

This paper is organized as follows. Section~\ref{sec:observations} summarizes our observations and reduction methods for both ground-based and space-based data; Section~\ref{sec:sample} describes our selection of the raw sample; in Section~\ref{sec:methods} we provide the methodology for analyzing individual events, including four-fold solutions, distance and mass estimations, and planet sensitivity computation. This method is then applied to the current sample, and results are presented in Section~\ref{sec:results}. In Section~\ref{sec:discussion} we discuss the implications of this work, as well as outline the path for future work.

\begin{deluxetable*}{lllrlccll}
\tablecaption{Summary of the 50 events in our sample. Here (RA,Dec) are the equatorial coordinates, and ($l$,$b$) are the Galactic coordinates. We also include the subjective selection dates and objective selection dates (if objective criteria are met), OGLE-IV Bulge fields and cadences. In the last column, we present the HJD dates of the first and last \emph{Spitzer} observation, as well as the total number of observations from \emph{Spitzer}.
\label{tab:evt-info}}
\tabletypesize{\scriptsize}
\tablehead{
\colhead{OGLE \#} & \colhead{RA (deg)} &
\colhead{Dec (deg)} & \colhead{$l$ (deg)} &
\colhead{$b$ (deg)} & \colhead{Subjective} &
    \colhead{Objective} & \colhead{OGLE-IV fields,} & \colhead{\emph{Spitzer} observations} \\
\colhead{} & \colhead{} & \colhead{} & 
\colhead{} & \colhead{} & \colhead{selection} &
\colhead{selection} & \colhead{cadences (per day)} & \colhead{start, stop, \#}
}
\startdata
0011 & 269.217833 & $-29.283250$ & $0.957550$ & $-2.289872$ & 5-30-11:59 & --- & BLG505, 30 & 7184.96, 7222.58, 53 \\
0029 & 269.944167 & $-28.644944$ & $1.828415$ & $-2.522683$ & 5-10-14:33 & 6-01 & BLG505, 30 & 7185.31, 7222.89, 52 \\
0034 & 270.580333 & $-27.516083$ & $3.088357$ & $-2.452817$ & 4-28-17:01 & 6-08 & BLG511, 10 & 7186.01, 7222.92, 62 \\
0081 & 268.653000 & $-28.996278$ & $0.957513$ & $-1.719135$ & 6-01-14:25 & --- & BLG505, 30 & 7184.10, 7221.81, 57 \\
0350 & 268.248583 & $-31.820278$ & $-1.657210$ & $-2.846465$ & 5-19-20:45 & 6-01 & BLG535, 3 & 7183.95, 7221.76, 61 \\
0379 & 269.104292 & $-29.574056$ & $0.656089$ & $-2.350021$ & 5-19-20:45 & 6-01 & BLG505, 30 & 7184.61, 7222.58, 54 \\
0388 & 268.468917 & $-28.534028$ & $1.274824$ & $-1.346174$ & 5-10-14:33 & 6-01 & BLG500, 10 & 7183.98, 7221.81, 66 \\
0461 & 270.043208 & $-28.156944$ & $2.295828$ & $-2.356385$ & 5-19-20:45 & --- & BLG504, 10 & 7185.79, 7222.90, 58 \\
0529 & 270.264667 & $-29.922917$ & $0.854524$ & $-3.397576$ & 5-16-22:18 & 6-08 & BLG513, 3 & 7185.79, 7222.92, 51 \\
0565 & 269.153708 & $-29.128056$ & $1.063892$ & $-2.163672$ & 5-16-22:18 & 6-01 & BLG505, 30 & 7184.62, 7222.59, 53 \\

\enddata
\tablecomments{This table is available in its entirety in the machine-readable format.}
\end{deluxetable*}

\section{Observations \& Data Reductions} \label{sec:observations}

\subsection{OGLE}
All events in our sample were found by the Optical Gravitational Lensing Experiment (OGLE) collaboration in real-time through its Early Warning System \citep{Udalski:1994,Udalski:2003}, based on observations with the 1.4~deg$^2$ camera on its 1.3-m Warsaw Telescope at the Las Campanas Observatory in Chile \citep{Udalski:2003,Udalski:2015b}. These events received OGLE-IV observations with cadences varying from 3 to 30 per day. The coordinates, OGLE-IV fields and cadences of individual events are provided in Table~\ref{tab:evt-info}.

OGLE data were reduced using the photometry software developed by \citet{Wozniak:2000} and \citet{Udalski:2003}, which was based on the Difference Image Analysis (DIA) technique \citep{AlardLupton:1998}.

\subsection{KMTNet}
The KMTNet consists of three 1.6-m telescopes located at CTIO in Chile, SAAO in South Africa, and SSO in Australia. Observations were initiated on 
February 3rd (JD=2457056.9), February 19th (JD=2457072.6), and June 9th (JD=2457182.9) in 2015 from CTIO, SAAO, and SSO, respectively. Each telescope is equipped with a 4 deg$^2$ field-of-view camera, and observes the $\sim$16 deg$^2$ prime microlensing fields at $\sim$10 min cadence when the bulge is visible.

The KMTNet data were reduced by the DIA photometric pipeline \citep{AlardLupton:1998,Albrow:2009}.

\subsection{\emph{Spitzer}} \label{sec:spitzer-obs}

As detailed in \citet{Yee:2015b}, the \emph{Spitzer} program is designed to maximize the sum of the products $\sum_i S_i P_i$, where $S_i$ is the planet sensitivity of event $i$ and $P_i$ is the probability to measure the microlens parallax of this event. As a consequence, the \emph{Spitzer} team started selecting targets beginning in early May, 2015, although \emph{Spitzer} did not start taking data until JD$'$=JD-2450000=7180.2 (2015 June 6.7).
To enforce our blindness to the existence of planets in any events, we select events if (1) they meet certain objective criteria at the time of one of the uploads of targets to \emph{Spitzer}, in which case they are considered as ``objectively chosen'', or (2) they do not meet objective criteria, but are nevertheless selected on the basis that the \emph{Spitzer} team believes that by selecting them the quantity $\sum_i S_i P_i$ can be maximized. Events selected in the latter case are known as ``subjectively chosen''. For objectively chosen events, planets as well as planet sensitivities from before or after the \emph{Spitzer} selection dates can be incorporated into the statistical analysis, while for subjectively chosen events, only planets (and planet sensitivities) from after the \emph{Spitzer} selection dates can be included in the final sample.
\footnote{More precisely, planets (and the putative planets needed for the sensitivity calculation) that are detectable in data that were available to the team prior to their decision, must be excluded.}
One relevant point is that, any event that is originally subjectively chosen but later meets objective criteria will be considered as objective chosen (provided its parallax is measurable based on the restricted set of \emph{Spitzer} data acquired after the date it became objective).

Events once selected are given \emph{Spitzer} cadences according to suggestions in \citet{Yee:2015b}. The majority of events received \emph{Spitzer} observations at 1/day cadence. Higher cadences were assigned to a few events, if the \emph{Spitzer} team believed the nominal cadence would lead to failures in parallax measurements. After all targets were scheduled according to their adopted cadences, the remaining time, if any, was applied to events that appeared or would appear with relatively high-magnification as seen from the ground. \emph{Spitzer} observations stopped if the pre-defined criteria for stopping observations in \citet{Yee:2015b} were met, or the event exited the \emph{Spitzer} Sun-angle window. Our last \emph{Spitzer} observation was taken on JD$'=7222.28$. In Table~\ref{tab:evt-info} we provide the information of \emph{Spitzer} selection and observation of each individual event.

\emph{Spitzer} data were reduced using the customized software that was developed by \citet{SCN:2015b} specifically for this program. This software improved the performance of \emph{Spitzer} IRAC photometry in crowded fields, although unknown systematics may persist in some cases. We discuss this in Section~\ref{sec:systematics}.

\subsection{Additional Color Data}
The characterization of a microlensing event requires a measurement of the color of the source star. This is usually achieved by using the less frequent $V$ band observations from survey teams, but it does not work for events that are highly extincted in optical bands. For this reason, we also obtained observations of all \emph{Spitzer} targets using the ANDICAM \citep{Depoy:2003} dichroic camera on the 1.3~m SMARTS telescope at CTIO. These observations were made simultaneously in $I$ and $H$ bands, and were for the specific purpose of inferring the $I-[3.6\microm]$ color of the source star. These additional color data were reduced using DoPhot \citep{Schechter:1993}.

\section{Raw Sample Selection} \label{sec:sample}

According to \citet{Yee:2015b}, only events in which $\bdv{\pi_\e}$ can be ``measured'' are useful for the study of the Galactic distribution of planets. While the phrase ``$\bdv{\pi_\e}$ is measured'' is not defined until Section~\ref{sec:measure-pie}, we provide here our procedure for raw sample selection.

In 2015, there are in total 68 \emph{Spitzer} events that fall within the footprints of KMTNet prime fields. The following events are excluded from the raw sample for various reasons:
\begin{itemize}
    \item[-] Three were not covered by OGLE; they were selected for \emph{Spitzer} observations based on alerts by MOA: MOA-2015-BLG-079, MOA-2015-BLG-237, and MOA-2015-BLG-267.
    \item[-] Event OGLE-2014-BLG-0613 was alerted in 2014; it has extremely long timescale and has not reached baseline by the time this study started.
    \item[-] Event OGLE-2015-BLG-1136 was later on identified as a cataclysmic variable (CV) rather than a microlensing event.
    \item[-] Six events show perturbations that can only be explained by stellar binaries: OGLE-2015-BLG-(0060, 0914, 0968, 1346, 1368) and OGLE-2015-BLG-1212 \citep{Bozza:2016}.
    \item[-] Events OGLE-2015-BLG-0022 and OGLE-2015-BLG-0244 show significant contamination of xallarap effect (binary-source orbital motion).
    \item[-] Events OGLE-2015-BLG-1109 and OGLE-2015-BLG-1187 have impact parameters as seen from Earth $u_{0,\oplus}>1$, which implies extremely low planet sensitivities.
    \item[-] The microlens parallax vector $\bdv{\pi_\e}$ of events OGLE-2015-BLG-1184 and OGLE-2015-BLG-1500 could not be measured, because the time coverages by \emph{Spitzer} are too short and the \emph{Spitzer} light curves do not show any features of microlensing \citep{SCN:2015b}. 
    \item[-] The microlens parallax vector $\bdv{\pi_\e}$ of event OGLE-2015-BLG-1403 could not be constrained because of the lack of the source color constraint.
\end{itemize}
Therefore, our raw sample contains 50 events. Information regarding their (equatorial and Galactic) positions and observations (by OGLE and Spitzer) is given in Table~\ref{tab:evt-info}. Since all of these events lie in one of the four prime KMTNet fields, which were observed essentially continuously, their KMTNet cadences are virtually identical.

\section{Methods} \label{sec:methods}
The sensitivity to planets of a microlensing event with a parallax measurement (and hence of an ensemble of such events) can be logically divided into two distinct problems. First, one must determine the probability function of the lens ``distance'' (defined more precisely below). Second, for each allowed distance, one must determine the sensitivity to planets as a function of planet parameters, either the microlensing $(q,s)$ or the physical parameters $(m_p,a_\perp)$. These issues have been previously addressed separately by \citet{SCN:2015a}, \citet{Yee:2015b}, and \citet{Zhu:2015b}. However, since this is the first measurement of sensitivity to the Galactic distribution of planets, we likewise present here the first integrated overview of the mathematics of this measurement. Moreover, based on this integration, we will identify some previously overlooked components of the analysis and also modify some past procedures.

Descriptions of the derivation of event solutions (Section~\ref{sec:solutions}), the estimation of lens distance and mass distributions (Section~\ref{sec:bayes}), and the computation of planet sensitivities (Section~\ref{sec:sensitivity}) follow immediately below.

\subsection{Four-Fold Solutions} \label{sec:solutions}
The separation between Earth and the satellite perpendicular to the line of sight to the microlensing event, $D_\perp$, causes apparent changes in the angular lens-source separation $\Delta \theta=\pi_\rel D_\perp/\au$, and this in turn gives rise to different microlensing light curves. These light curves, as seen from Earth and from the satellite, appear to peak at different times $t_0$ and with different impact parameters $u_0$ (normalized to $\theta_\e$). In the approximation of rectilinear motion of Earth and the satellite \citep{Refsdal:1966,Gould:1994,Graff:2002}
\begin{equation} \label{eqn:satpie0}
    \bdv{\pi_\e} \approx \frac{\au}{D_\perp} \left(\Delta\tau,\Delta\beta\right)\ ,
\end{equation}
where
\begin{equation}
    \Delta\tau \equiv \frac{t_{0,\sat}-t_{0,\oplus}}{t_\e};\quad
    \Delta\beta \equiv u_{0,\sat}-u_{0,\oplus}\ .
\end{equation}
Unfortunately, $u_0$ is a signed quantity (depending on whether the lens passes the source on its right or left, see Fig.~4 of \citealt{Gould:2004} for sign definition), while only $|u_0|$ is directly measurable from the light curve.  Therefore, satellite parallax measurements are subject to a four-fold degeneracy
\footnote{Here we adopt the following notation for the degenerate solutions: $[{\rm sgn}(u_{0,\oplus}),{\rm sgn}(u_{0,\rm spitz})]$. See \citet{Zhu:2015a} for the conversion between this notation and the one used in \citet{SCN:2015a}.}
\begin{equation} \label{eqn:satpie}
    \bdv{\pi_\e} \approx \frac{\au}{D_\perp} \left(\Delta\tau,\Delta\beta_{\pm,\pm}\right)\ ,
\end{equation}
where
\begin{equation}
    \left\{
        \begin{array}{ll}
            \Delta\beta_{+,+} \equiv+u_{0,\sat}-u_{0,\oplus} & ,(+,+){\rm\ solution} \cr
            \Delta\beta_{+,-} \equiv-u_{0,\sat}-u_{0,\oplus} & ,(+,-){\rm\ solution} \cr
            \Delta\beta_{-,-} \equiv-u_{0,\sat}+u_{0,\oplus} & ,(-,-){\rm\ solution} \cr
            \Delta\beta_{-,+} \equiv+u_{0,\sat}-u_{0,\oplus} & ,(-,+){\rm\ solution} \cr
        \end{array}\right.\ .
\end{equation}

In principle, higher-order effects in the light curve itself can break this degeneracy. At first order (in the polynomial expansion of \citealt{Smith:2003}), it can be broken from the different Einstein timescales $t_\e$ measured from Earth and satellite due to their relative motion (even within the approximation of rectilinear motion) \citep{Gould:1995}. At third and fourth order, it can be broken due to parallax effects from the accelerated motion of Earth \citep{Gould:1992}. In practice, however, these effects are usually quite weak. First, with current experiments, $t_\e$ is normally not independently measured from the satellite simply because the observational investment for this would be extremely high \citep{GaudiGould:1997a}, and these resources are better applied to observing more events. Ground-based parallaxes are rarely measured because the Einstein timescales are typically small $t_\e < {\rm yr}/2\pi$, so that third, and particularly fourth, order effects are very subtle. This indeed is the reason for going to space. Nevertheless, although these higher-order effects are small, they can contribute to breaking the degeneracy between well-determined, but otherwise indistinguishable parallax solutions.

We search for and characterize the four solutions using the code developed in \citet{Zhu:2016a}. We first find a simple three-parameter $(t_0,u_0,t_\e)$ solution based on OGLE data. Next, we include \emph{Spitzer} data, introduce two parameters $\pi_\en$ and $\pi_\ee$, which are the two components of vector $\bdv{\pi_\e}$ along the north and east directions, respectively, and easily find one of the four parallax solutions by allowing $\chi^2$ to go downhill. As per the usual convention, these parameters ($\pi_\en$, $\pi_\ee$, $t_\e$) are defined in the geocentric frame \citep{Gould:2004}.
\footnote{See a discussion of microlensing parallax in heliocentric frame in \citet{SCN:2016}.}
The location of the \emph{Spitzer} satellite is extracted from the \emph{JPL Horizons} website \footnote{\url{http://ssd.jpl.nasa.gov/?horizons}}, enabling a self-consistent quantification of the microlens parallax effect and the event timescale. 
In addition, there are two flux parameters for each data set, $F_{\rm s}$ and $F_{\rm b}$. The former is the flux from the source, and the latter is the flux that is blended within the aperture and does not participate in the event. The model for the total flux observed at epoch $t_i$ for data set $j$ is then given by
\begin{equation}
    F^j(t_i) = F_{\rm s}^j \cdot A^j(t_i;t_0,u_0,t_\e,\rho,\bdv{\pi_\e}) + F_{\rm b}^j\ .
\end{equation}
Once a solution is found, we estimate the uncertainties of parameters via a Markov Chain Monte Carlo (MCMC) analysis, using the \texttt{emcee} ensemble sampler \citep{emcee:2013}. The remaining three solutions are also easily found by seeding solutions at the locations expected based on Equation~(\ref{eqn:satpie}). In some cases, typically events with long timescales or events peaking near the beginning of the season, there is no local minimum at $\chi^2$ surface for one or more solutions due to strong parallax information from the ground. Within the mathematical formalism that follows, these other solutions can be thought of as ``existing'' but having very high $\Delta \chi^2$ relative to the best solution. 

For each of the four solutions we then derive $\bdv{\pi_\e}$, $\bdv{\tilde{v}_\hel}$, the uncertainty of the latter quantity, and $\Delta \chi^2$ relative to the best solution. Here, $\bdv{\tilde{v}_\hel}$ is the transverse velocity between the source and the lens projected onto the observer plane, after the correction from geocentric to heliocentric frames,
\begin{equation} \label{eqn:vvec}
    \bdv{\tilde{v}_\hel} = \bdv{\tilde{v}_\geo} + \bdv{v}_{\oplus,\perp};\quad
    \bdv{\tilde{v}_\geo} = \frac{\au}{t_\e}\frac{\bdv{\pi_\e}}{\pi_\e^2}\ ,
\end{equation}
where $\bdv{v}_{\oplus,\perp}$ is the velocity of Earth at the event peak and projected perpendicular to the directory of the event. To facilitate further discussions, we also define here the event timescale in the heliocentric frame
\begin{equation}
    t_\e^\prime \equiv \frac{\tilde{r}_\e}{\tilde{v}_\hel};\quad \tilde{r}_\e \equiv \frac{\au}{\pi_\e}\ .
\end{equation}

In deriving event solutions, we are able to incorporate ``color constraints'' (either $VI[3.6\,\microm]$ or $IH[3.6\,\microm]$) into the fit. This is either very important or essential for the great majority of cases, as anticipated by \citet{Yee:2015b}. The naive idea of space-based parallaxes, as outlined by \citet{Refsdal:1966} and \citet{Gould:1994} and as captured by Equation~(\ref{eqn:satpie0}), is that $t_0$ and $u_0$ will be measured independently from the satellite and Earth. However, such independent measurements are essentially impossible if the event is not observed over (or at least close to) peak. Hence, in the 2014 pilot program, exceptional efforts were made to observe over peak, which greatly restricted the number of events that could be targeted, given the short ($\sim$38 day) observing window set by \emph{Spitzer} Sun-angle restrictions and given the $6\pm3$ day delays in observing targets (Fig.~1 of \citealt{Udalski:2015a}).  However, based on experience in optical bands \citep{Yee:2012}, \citet{Yee:2015b} argued that, even if the peak were not observed from the satellite, it would be possible to recover $(t_0,t_\e)_\sat$ provided that the {\it Spitzer} source flux could be determined from a combination of (1) the measured source flux of the ground-based light curve, (2) the measured source color in ground-based bands ($V-I$ or $I-H$), and (3) a color-color relation (e.g., $VI[3.6\,\microm]$) derived from field stars. In practice, we derive the $(I-[3.6\,\microm])$ from the measured color and color-color relation and then impose the $2\,\sigma$ limits of this measurement as hard constraints in the fit.

\begin{figure*}[ht]
    \epsscale{1.2}
    \plotone{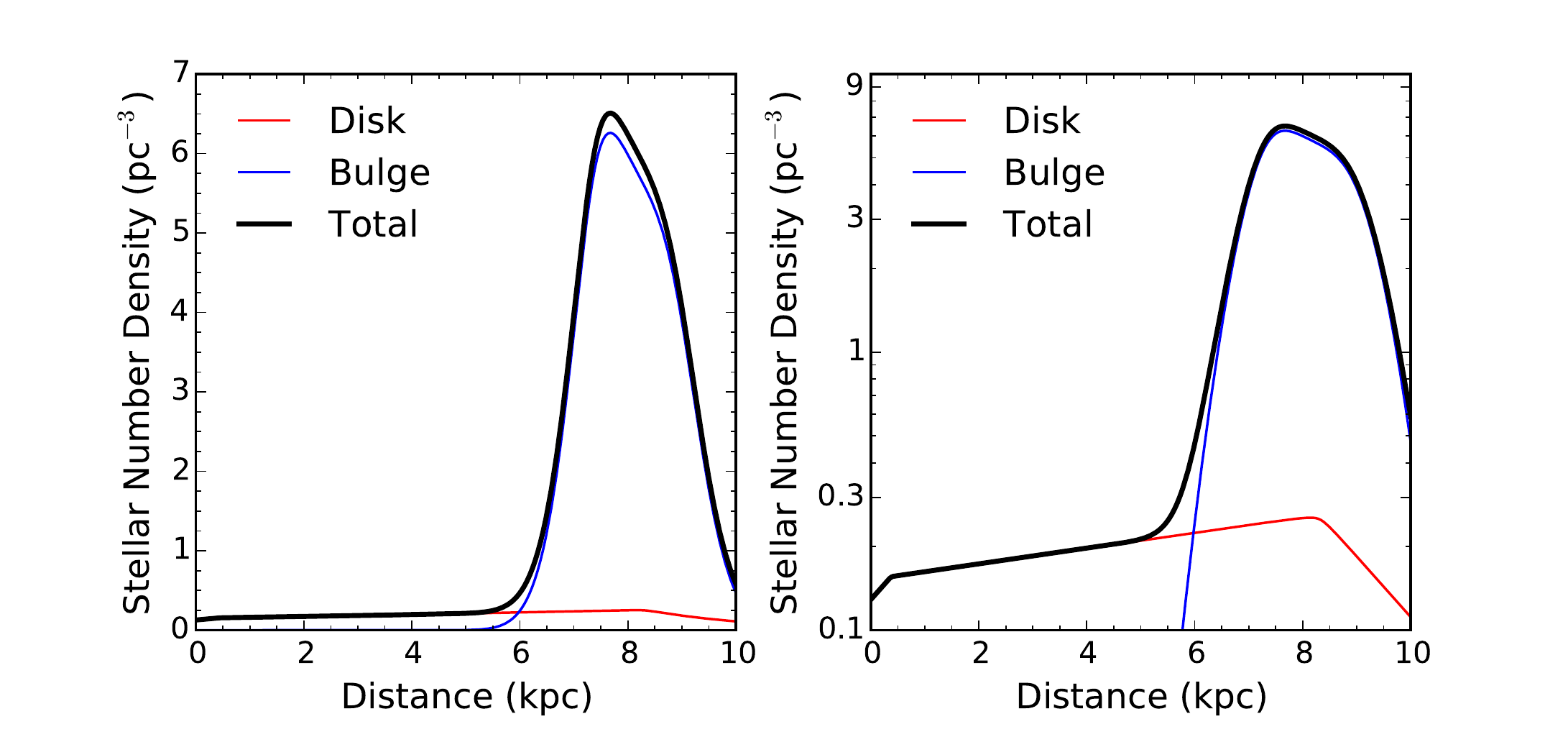}
    \caption{The stellar number density profile toward the Baade's window for our adopted Galactic model, shown in linear scale on the left and logarithmic scale on the right.
    \label{fig:density-profile}}
\end{figure*}

\subsection{Galactic Model} \label{sec:galmod}

\subsubsection{Stellar Density Profile}
The Galactic Center has equatorial coordinates $(\alpha_{\rm GC},\delta_{\rm GC})=(17^{\rm h}45^{\rm m}37\fs224,-28\arcdeg56\arcmin10\farcs23)$ \citep{Reid:2004} and heliocentric distance $\Rgc=8.3~\kpc$ \citep{Gillessen:2009}. The Sun is above the Galactic mid-plane $(z=0)$ by $27~$pc \citep{Chen:2001}, which corresponds to a tilt angle $\beta=0.19^\circ$.

The total stellar number density $n_\star$ at given Galactocentric coordinates $(x,y,z)$ is the sum of contributions from the bulge and disk components
\begin{equation} \label{eqn:ntot}
    n_{\star}(x,y,z) = n_{\rm B}(x^\prime,y^\prime,z^\prime) + n_{\rm D} (R,z)\ .
\end{equation}
We assume a triaxial G2 model for the bulge component \citep{Kent:1991,Dwek:1995}.
\begin{equation}
\small
    n_{\rm B} = n_{\rm B,0} e^{-r_s^2/2};\quad 
    r_s \equiv \left\{\left[\left(\frac{x'}{x_0}\right)^2 + \left(\frac{y'}{y_0}\right)^2\right]^2 + \left(\frac{z'}{z_0}\right)^4\right\}^{1/4}\ ,
\end{equation}
where $n_{\rm B,0}=13.7$~pc$^{-3}$, $x_0=1.59~$kpc, $y_0=424~$pc, and $z_0=424~$pc. These values are adopted from \citet{Robin:2003}. The coordinates $(x',y',z')$ are derived by rotating the Galactocentric coordinates $(x,y,z)$ around $z$ axis by $\alpha_{\rm bar} = 30^\circ$ \citep[e.g.,][]{Cao:2013,WeggGerhard:2013}. The disk component in Equation~(\ref{eqn:ntot}) has the form \citep{Bahcall:1986}
\begin{equation}
    n_{\rm D} = n_{\rm D,0} \exp \left[-\left(\frac{R-R_{\rm GC}}{R_0}+\frac{|z|}{z_{\rm D,0}}\right)\right]\ .
\end{equation}
Here $R\equiv \sqrt{x^2+y^2}$, the local stellar number density $n_{\rm D,0}=0.14$~pc$^{-3}$, the scale length of the disk $R_0=3.5~\kpc$, and the scale height of the disk $z_{\rm D,0}=325~$pc \citep{Han:1995}. We show in Figure~\ref{fig:density-profile} the stellar number density profile toward the Baade's window, which is approximately the center of microlensing fields. 

\subsubsection{Stellar Velocity Distribution}

The mean stellar velocity at Galactocentric coordinates $(x,y,z)$ has the form
\begin{equation}
    \bdv{\mu_v}(x,y,z) = \frac{n_{\rm B}}{n_\star}\bdv{\mu}_{v,\rm B} + \frac{n_{\rm D}}{n_\star}\bdv{\mu}_{v,\rm D}\ ,
\end{equation}
and the velocity dispersion is given by
\begin{equation}
    \sigma_{v,i}^2(x,y,z) = \left(\frac{n_{\rm B}}{n_\star}\right)^2{\sigma}^2_{v,i,\rm B} + \left(\frac{n_{\rm D}}{n_\star}\right)^2{\sigma}^2_{v,i,\rm D}\quad (i=x,y,z)\ .
\end{equation}
We assume that the bulge stars have zero mean velocity and $120~\kms$ velocity dispersion along each direction ($\sigma_{v,i,\rm B}=120~\kms$). The latter is derived from the proper motion dispersion of bulge stars $\sigma_{\mu}=3~\masyr$ \citep{Poleski:2013}. Disk stars partake of the flat rotation curve with $240~\kms$ (i.e., $\mu_{v,z,\rm D}=0~\kms$, and $\mu_{v,y,\rm D}=240~\kms$, \citealt{Reid:2014}), and their velocity dispersions are $18~\kms$ and $33~\kms$ in the vertical $(z)$ and rotation $(y)$ directions. The Sun partakes of the same rotation curve, and has a peculiar motion ($V_\odot=12~\kms$ and $W_\odot=7~\kms$, \citealt{Schoenrich:2010}) relative to the local standard of rest.

\subsubsection{Stellar Mass Function}
We choose two forms of the lens mass function (MF): (1) a flat MF with $\dif\xi(M_\l)/\dif\log{M_\l}\propto 1$; and (2) a Kroupa MF \citep{Kroupa:2001}
\begin{equation} \label{eqn:kroupa}
\frac{\dif\xi(M_\l)}{\dif\log{M_\l}} \propto \left\{ 
    \begin{array}{ll}
    M_\l^{0.7} &,\ 0.013<M_\l/M_\odot<0.08 \cr
    M_\l^{-0.3} &,\ 0.08<M_\l/M_\odot<0.5 \cr
    M_\l^{-1.3} &,\ 0.5<M_\l/M_\odot<1.3
\end{array} \right.\ .
\end{equation}
In both cases, no planetary lenses are included, and the upper end of the MF is truncated at $1.3M_\odot$. As has been demonstrated in \citet{SCN:2015a} and will also be shown later, the choice of a different MF has essentially no effect on the result.

\begin{figure*}[bt]
    \epsscale{1.2}
    \plotone{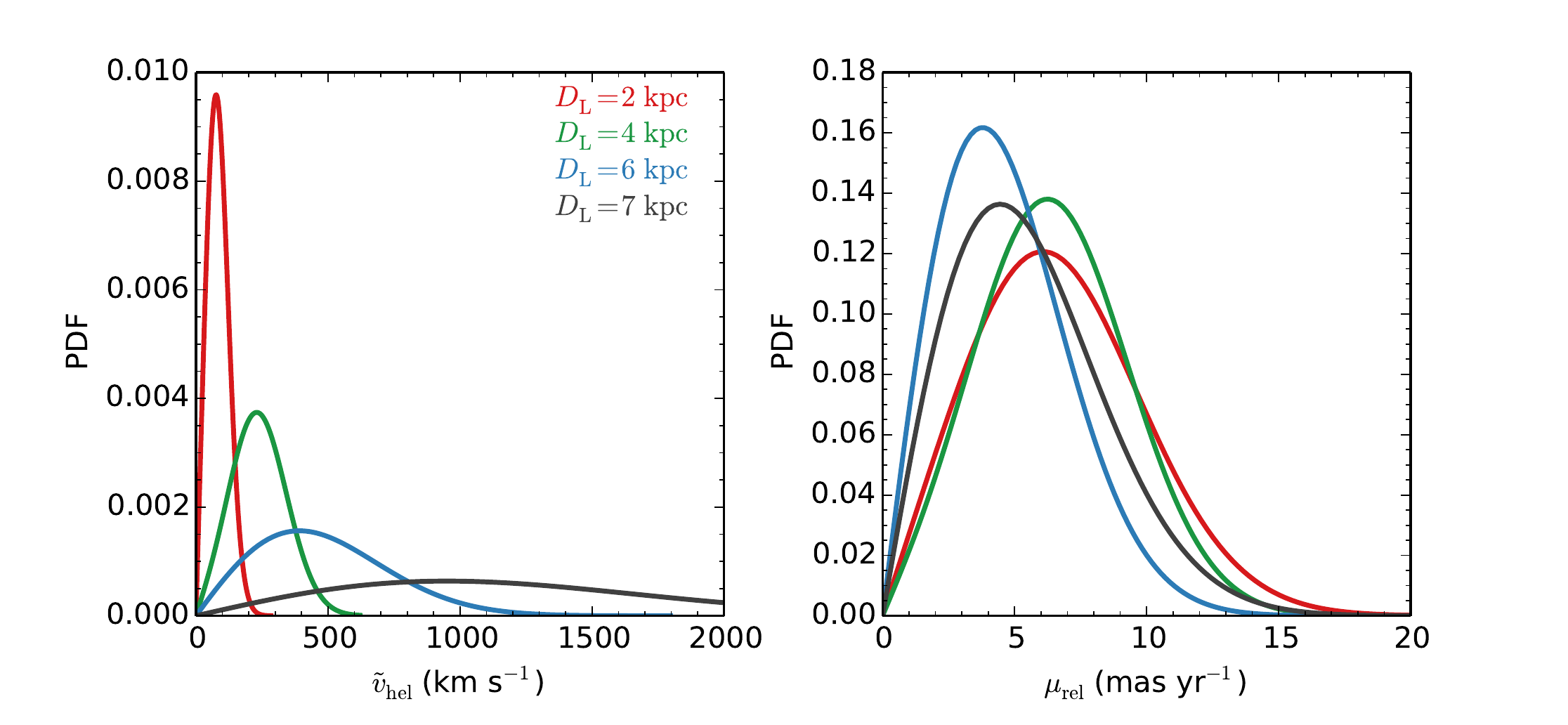}
    \caption{The (prior) probability distributions of $\tilde{v}_\hel$ (left panel) and $\mu_\rel$ (right panel) for four lens distances $D_\l=2$, 4, 6, and $7~\kpc$, under the Galactic model specified in Section~\ref{sec:galmod}. These distances represent typical lens distances at near disk, mid-disk, far disk, and bulge, respectively. For this illustration, the source has a fixed distance $D_\s=8.3~\kpc$.
    \label{fig:prior}}
\end{figure*}

\subsection{The Lens Distance \& Mass Distribution} \label{sec:bayes}

Following \citet{SCN:2015a}, we define a lens ``distance'' parameter $\Dparm$ that is a monotonic function of $\pi_\rel$,
\begin{equation} \label{eqn:ddef}
    \Dparm \equiv \frac{\kpc}{\left(\pi_\rel/\mas\right) + 1/8.3}\ .
\end{equation}
This has the advantage that $\pi_\rel$ is much better constrained than the lens distance $D_\l$ (see Equation~\ref{eqn:mass}), and also informs us more of the Galactic population from which the lens is drawn. That is,
\begin{equation}
    \Dparm \rightarrow D_\l \quad (D_\l \ll D_\s)\ ;
\end{equation}
\begin{equation}
    (8.3~\kpc-\Dparm) \rightarrow D_{\l\s} \quad (D_{\l\s}\ll D_\s)\ ,
\end{equation}
where $D_{\l\s}\equiv D_\s-D_\l$. A determination that $D_{\l\s} \ll D_\s$ is a much better indicator that the lens is in the bulge than the value of $D_\l$ (which in any case is less precisely known).

As discussed in Section~\ref{sec:introduction}, the lens distance parameter $\Dparm$ cannot be uniquely determined for the majority of events because of the lack of $\theta_\e$ measurement. We therefore derive the Bayesian distribution of $\Dparm$ by imposing a Galactic model.  As first pointed out by \citet{Han:1995}, such a distribution of $\Dparm$ is fairly compact if $\bdv{\pi_\e}$ rather than $\theta_\e$ (which gives $\mu_\rel$) can be measured. One can understand this by first approximating the Galactic disk lenses as moving exactly on a flat rotation curve and bulge sources as not moving. Then (also approximating the Sun as being at the local standard of rest),
\begin{equation}
    \pi_\rel \rightarrow \frac{\tilde{v}_\hel}{\mu_{\rm sgrA\star}}\ ,
\end{equation}
where $\mu_{\rm sgrA\star}$ is the observed proper motion of the Galactic center, and $\tilde{v}_\hel$ is the magnitude of $\bdv{\tilde{v}_\hel}$ from Equation~(\ref{eqn:vvec}). In fact, the velocities of both the sources and lenses are dispersed relative to this naive model. However, since these dispersions (projected on the observer plane) are typically small compared to the projection of the flat rotation curve, the probability distribution of $\pi_\rel$ (and therefore $\Dparm$) is typically compact. Then, since $M_\l=\pi_\rel/(\kappa\pi_\e^2)$, $M_\l$ is also quite well measured. 

To further illustrate this point under our adopted Galactic model, we show in Figure~\ref{fig:prior} the probability distributions of $\tilde{v}_\hel$ and $\mu_\rel$ for several different lens distances. Here $\tilde{v}_\hel$ and $\mu_\rel$ are the amplitude of the two vectors, $\bdv{\tilde{v}_\hel}$ and $\bdv{\mu_\rel}$, respectively, and these vectors are related to lens and source properties by
\begin{equation} \label{eqn:vvec-gc}
    \bdv{\tilde{v}_\hel} = \frac{D_\s}{D_{\l\s}} \bdv{v}_{\l,\gc} - \frac{D_\l}{D_{\l\s}} \bdv{v}_{\s,\gc} - \bdv{v}_{\odot,\gc}\ ,
\end{equation}
and
\begin{equation} \label{eqn:murel-gc}
    \bdv{\mu_\rel} = \frac{\bdv{v}_{\l,\gc}-\bdv{v}_{\odot,\gc}}{D_\l} - \frac{\bdv{v}_{\s,\gc}-\bdv{v}_{\odot,\gc}}{D_\s}\ .
\end{equation}
Here $\bdv{v}_{\l,\gc}$, $\bdv{v}_{\s,\gc}$, and $\bdv{v}_{\odot,\gc}$ are the Galactocentric velocities of the lens, the source, and Sun, respectively. Figure~\ref{fig:prior} demonstrates again that any knowledge of $\tilde{v}_\hel$ provides much more information of the lens distance than $\mu_\rel$ could be.

The distribution of $\Dparm$ is derived following a variant of the method in \citet{SCN:2015a}. Here we provide the mathematical form of this derivation. For a fixed source distance $D_\s$, the differential event rate of Galactic microlensing is given by
\begin{equation} \label{eqn:evt-rate}
        \frac{\dif^4\Gamma}{\dif D_\l \dif \log M_\l \dif^2 \bdv{\mu_\rel}} = n_{\l,\star} D_\l^2 (2\theta_\e) \mu_\rel f_{\mu}(\bdv{\mu_\rel})\frac{\dif\xi(M_\l)}{\dif \log M_\l}\ .
\end{equation}
Here $n_{\l,\star}$ is the local stellar density at position $(\alpha,\delta,D_\l)$, $f_{\mu}(\bdv{\mu_\rel})$ is the two-dimensional probability distribution function of the lens-source relative proper motion $\bdv{\mu_\rel}$, and $\dif\xi(M_\l)/\dif \log M_\l$ is the stellar mass function in logarithmic scale. Equation~(\ref{eqn:evt-rate}) can be rewritten in terms of microlensing observables $(\Dparm,t_\e^\prime,\bdv{\tilde{v}_\hel})$,
\begin{equation} \label{eqn:evt-rate2}
    \begin{aligned}
        \frac{\dif^4\Gamma}{\dif \Dparm \dif t_\e^\prime \dif^2\bdv{\tilde{v}_\hel}} & =  \frac{\dif^4\Gamma}{\dif D_\l \dif \log M_\l \dif^2 \bdv{\mu_\rel}} \left| \frac{\partial(D_\l,\log M_\l,\bdv{\mu_\rel})}{\partial(\Dparm,t_\e^\prime,\bdv{\tilde{v}_\hel})}\right| \cr
        & =  4n_{\l,\star} \frac{D_\l^4}{\Dparm^2} \mu_\rel^2 f_{\tilde{v}}(\bdv{\tilde{v}_\hel}) \frac{\dif \xi(M_\l)}{\dif \log M_\l}
    \end{aligned}\ .
\end{equation}
Here $f_{\tilde{v}}(\bdv{\tilde{v}_\hel})$ is the two-dimensional probability function of $\bdv{\tilde{v}_\hel}$, which can be derived from Equation~(\ref{eqn:vvec-gc}) under a given Galactic model. In the latter evaluation of Equation~(\ref{eqn:evt-rate2}), we have substituted Equation~(\ref{eqn:evt-rate}) and the following Jacobian determinant
\begin{equation}
    \begin{aligned}
        \left| \frac{\partial(D_\l,\log M_\l,\bdv{\mu_\rel})}{\partial(\Dparm,t_\e^\prime,\bdv{\tilde{v}_\hel})}\right| &= \left(\frac{D_\l}{\Dparm}\right)^2 \frac{\mu_\rel}{\tilde{v}_\hel}\frac{1}{M_\l} \left| \frac{\partial(M_\l,\mu_\rel)}{\partial(t_\e^\prime,\tilde{v}_\hel)}\right| \cr
        &= \left(\frac{D_\l}{\Dparm}\right)^2 \frac{2\pi_\rel^2}{\au^2 t_\e^\prime}
    \end{aligned}\ .
\end{equation}

For a given set of $(t_\e^\prime,\bdv{\tilde{v}_\hel})$, Equation~(\ref{eqn:evt-rate2}) thus determines the relative (prior) probability distribution of $\Dparm$ at fixed $D_\s$. This is then integrated over the posterior distributions of $t_\e^\prime$ and $\bdv{\tilde{v}_\hel}$ from the light curve modeling to yield the relative probability distribution of $\Dparm$ for a fixed $D_\s$.
To account for the variation in $D_\s$, we average over all possible values of $D_\s$ (from $D_\min=6~\kpc$ to $D_\max=10~\kpc$, assuming bulge sources), with each weighted by the number of available sources at that distance, $n_{\s,\star}D_\s^{2-\gamma}\dif D_\s$. Here $n_{\s,\star}$ is the local stellar density at $(\alpha,\delta,D_\s)$, $D_\s^2\dif D_\s$ is the volume between $D_\s$ and $D_\s+\dif D_\s$, and $D_\s^{-\gamma}$ is approximately the fraction of stars that have the measured apparent magnitude \citep{Kiraga:1994}. We choose $\gamma=2.85$ for our sample, for reasons that are given in Appendix~\ref{sec:gamma}. Then the non-normalized (``raw'') probability distribution of $\Dparm$ for the given solution is
\begin{equation} \label{eqn:ddist-prior}
    P_{\rm raw} (\Dparm) = \frac{\int_{D_{\s,\min}}^{D_{\s,\max}} \mathcal{P}(\Dparm|D_\s) n_{\s,\star} D_\s^{2-\gamma} \dif D_\s}{\int_{D_{\s,\min}}^{D_{\s,\max}} n_{\s,\star} D_\s^{2-\gamma} \dif D_\s}\ ,
\end{equation}
where
\begin{equation} \label{eqn:mathcalp}
    \small
    \mathcal{P}(\Dparm|D_\s) \equiv \int \frac{\dif^4\Gamma}{\dif \Dparm \dif t_\e^\prime \dif^2\bdv{\tilde{v}_\hel}} P(t_\e^\prime|\data)P(\bdv{\tilde{v}_\hel}|\data) \dif t_\e^\prime \dif^2 \bdv{\tilde{v}_\hel}\ .
\end{equation}
In practice, we assume that the posterior distribution of $t_\e^\prime$, $P(t_\e^\prime|\data)$, is a Dirac $\delta$ function, and that the posterior distribution of $\bdv{\tilde{v}_\hel}$, $P(\bdv{\tilde{v}_\hel}|\data)$, is a bivariate Gaussian function whose covariance matrix is determined in Section~\ref{sec:solutions}. The former assumption is reasonable because $t_\e^\prime$ (essentially $t_\e$) is well measured in almost all events and especially because it is much better constrained than $\bdv{\tilde{v}_\hel}$. The second assumption is adopted so that the above integration can be computed analytically (see Appendix~\ref{sec:gaussian}). We have nevertheless tested the validity of this assumption with some examples, by comparing the analytic result with numerical integration of the (non-Gaussian) true posterior distribution from MCMC.

To derive the distance distribution of one event, we must weight all degenerate solutions correctly. The weight contains two factors: (1) $\exp{(-\Delta\chi^2/2)}$, which is from the light curve modeling, and (2) $\pi_\e^{-2}$, which is based on the so-called ``Rich'' argument. The Rich argument was originally pointed out by James Rich (ca 1997, private communication). It argues that, qualitatively, if $\pi_{\e,+-}$ (and so $\pi_{\e,-+}$) are much larger than $\pi_{\e,++}$ (and $\pi_{\e,--}$), then the former are likely spurious solutions. This is because the true solutions for events with small $\pi_\e$ are much more likely to be $\pi_{\e,\pm\pm}$ solutions, and can almost always generate spurious counterpart solutions $\pi_{\e,\pm\mp}$ that are much larger. However, large $\pi_\e$ solutions can only rarely generate spurious small $\pi_\e$ solutions. \citet{SCN:2015a} quantified this argument and showed that solutions should be weighted by $\pi_\e^{-2}$, although they nevertheless only applied this weighting when the ratio of $\pi_\e$ between solutions was relatively large. Here, we carry the analysis of \citet{SCN:2015a} to its logical conclusion and apply this weighting uniformly to all events. The final normalized distribution of $\Dparm$ for each individual solution $i$ is given by
\begin{equation} \label{eqn:ddist-normed}
    P_{i}(\Dparm) = \frac{e^{-\Delta\chi_i^2/2}\pi_{\e,i}^{-2} P_{i,\rm raw}(\Dparm)}{\mathcal{N}},
\end{equation}
where
\begin{equation}
    \mathcal{N} \equiv \sum_{i=1}^{4} \frac{e^{-\Delta\chi_i^2/2}}{\pi_{\e,i}^2}\int P_{i,\rm raw}(\Dparm) \dif \Dparm\ .
\end{equation}

The lens mass distribution is derived in a similar way. One key equation involved is given below
\begin{equation}
    \frac{\dif^4\Gamma}{\dif \log M_\l \dif t_\e^\prime \dif^2 \bdv{\tilde{v}_\hel}} = 4n_{\l,\star} D_\l^4 f_{\tilde{v}}(\bdv{\tilde{v}_\hel}) \frac{\dif\xi(M_\l)}{\dif \log M_\l} \frac{\mu_\rel^3}{\tilde{v}_\hel}\ .
\end{equation}
The rest are almost identical to Equations~(\ref{eqn:ddist-prior}) and (\ref{eqn:ddist-normed}), except for replacing $\Dparm$ with $M_\l$ (or $\log M_\l$).

\begin{figure*}
    \centering
    \epsscale{1.2}
    \plotone{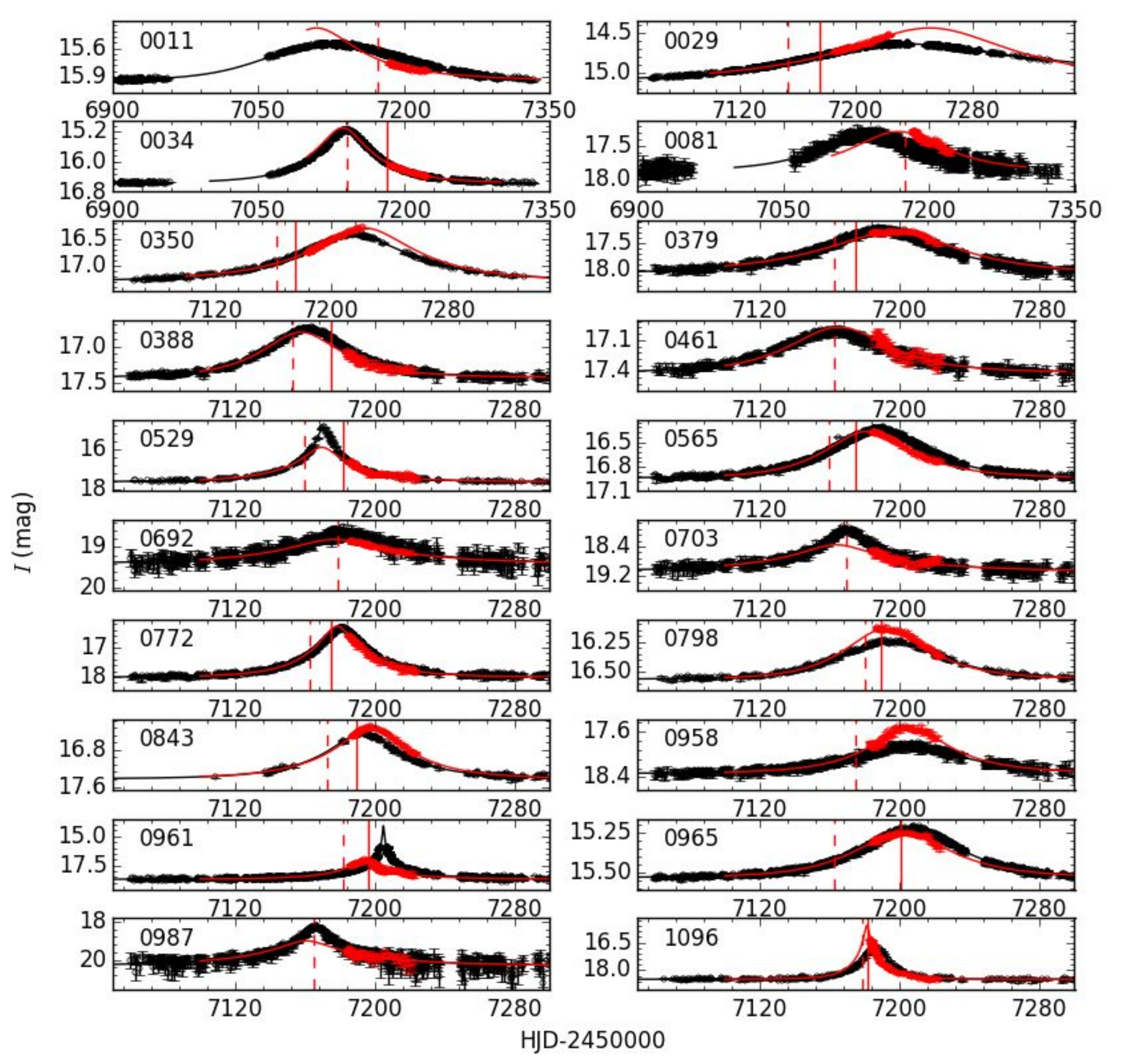}
    \caption{Ground-based (black) and space-based (red) data and best-fit models of the first 20 events in our sample. Here we only show the OGLE data for the ground-based part. KMTNet data have nearly continuous coverage with cadence $\sim$10 mins, as shown in Figure~\ref{fig:0961} for an example. The \emph{Spitzer} data and light curves have been rescaled to the OGLE magnitude system according to Equation~(\ref{eqn:rescaling}). For each event, the OGLE number is shown in the upper left, and the vertical red lines indicate the subjective (dashed) and objective (solid) selection dates. Note that models shown here are the ones with minimum $\chi^2$. Please refer to Table~\ref{tab:parameters} for parameters and uncertainties of individual events.
    \label{fig:lcs-1}}
\end{figure*}

\begin{figure*}
    \centering
    \epsscale{1.2}
    \plotone{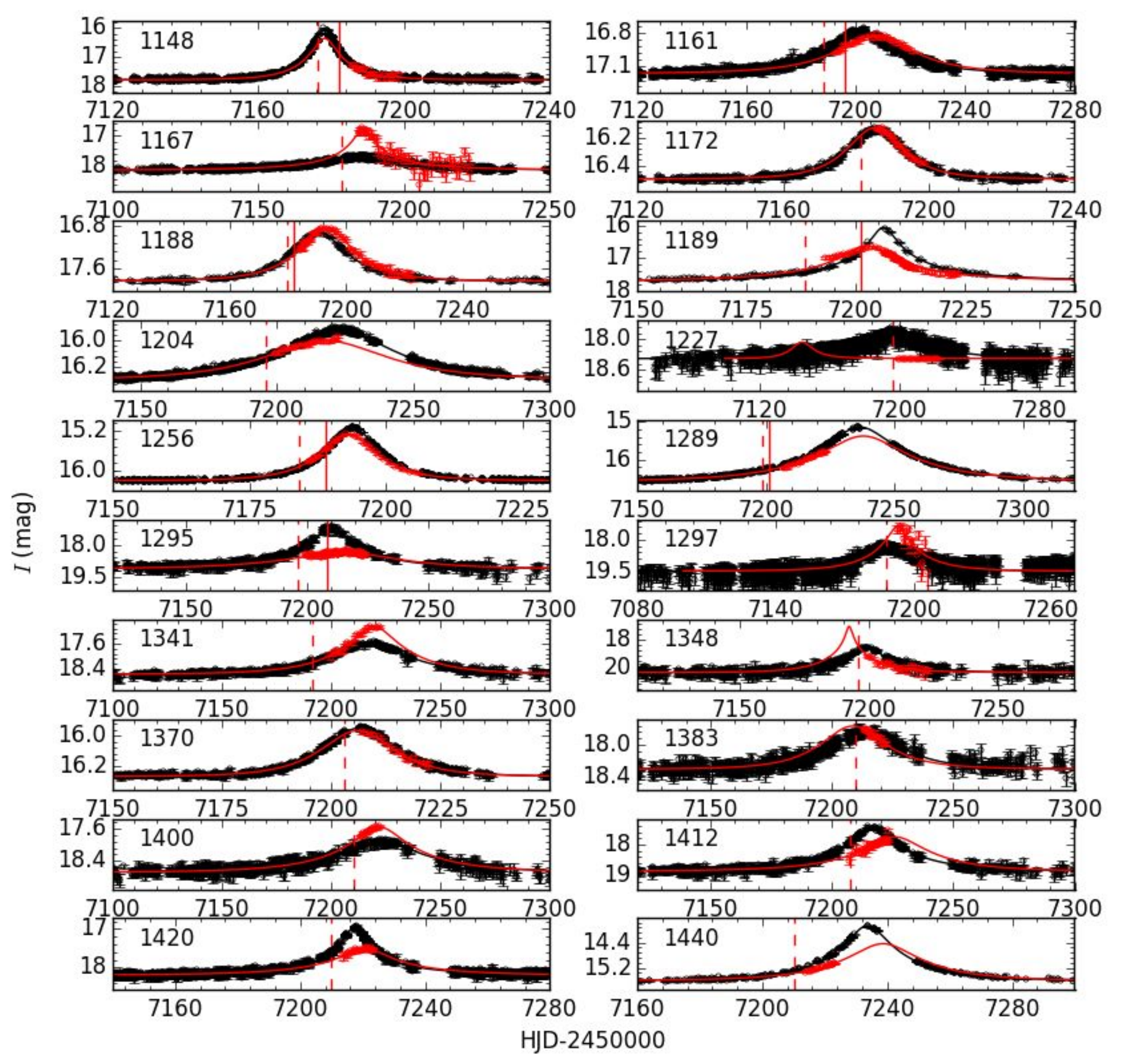}
    \caption{Ground-based (black) and space-based (red) data and best-fit models for events with OGLE number from 1148 to 1440. See Figure~\ref{fig:lcs-1} caption for detailed explanations.
    \label{fig:lcs-2}}
\end{figure*}

\begin{figure*}
    \centering
    \epsscale{1.2}
    \plotone{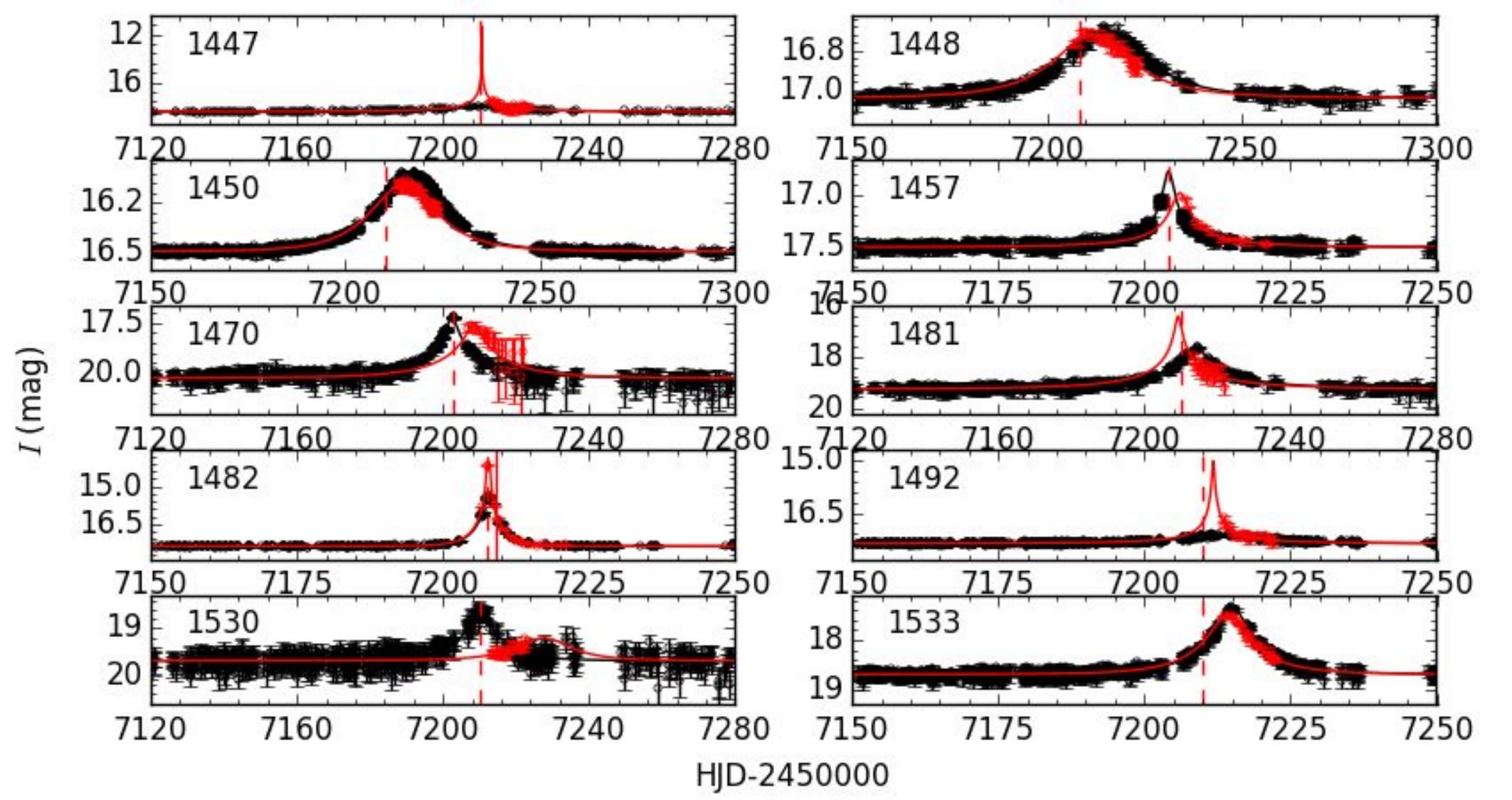}
    \caption{Ground-based (black) and space-based (red) data and best-fit models for the last 10 events in our sample. See Figure~\ref{fig:lcs-1} caption for detailed explanations.
    \label{fig:lcs-3}}
\end{figure*}

\begin{figure}
    \centering
    \epsscale{1.}
    \plotone{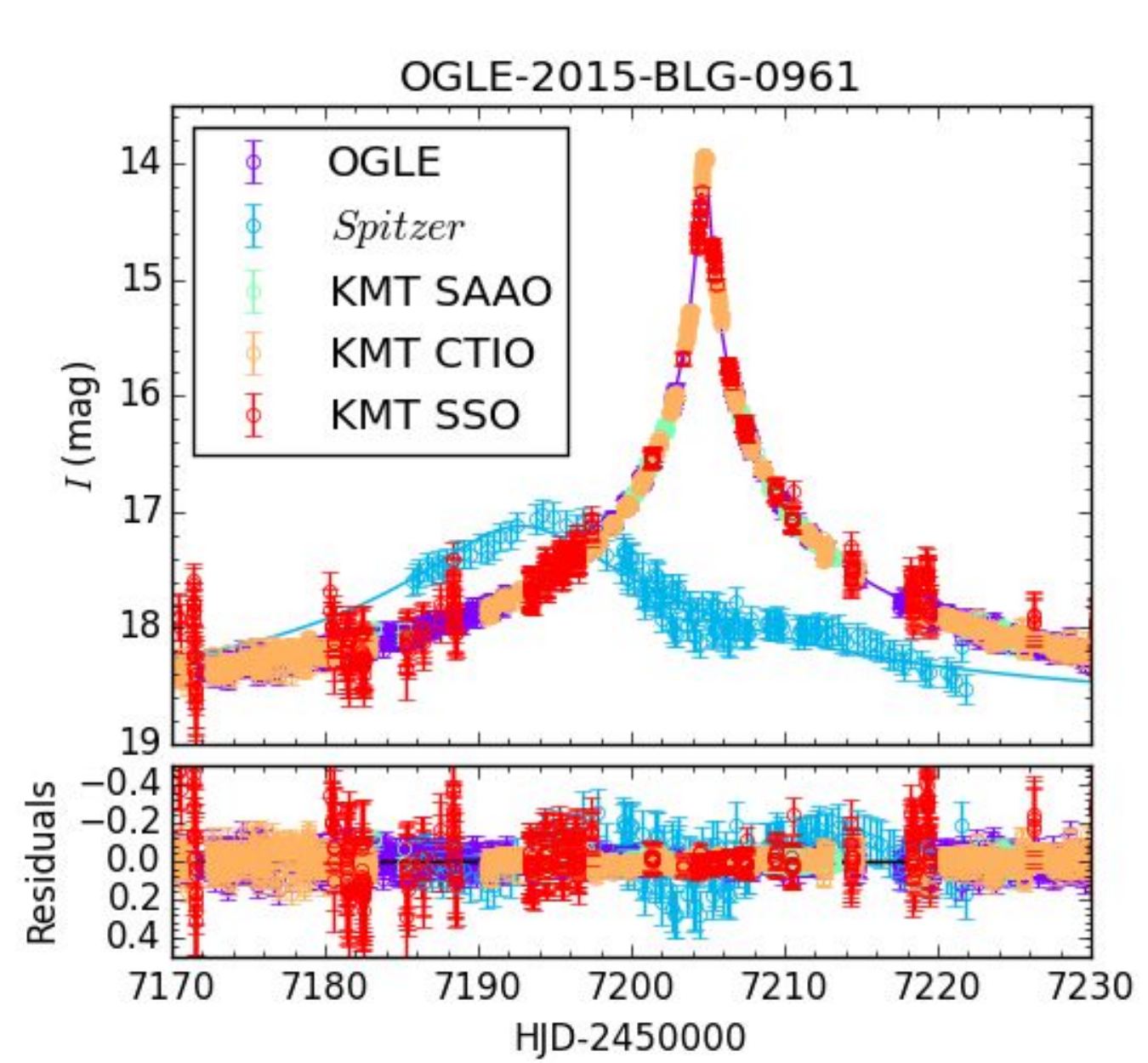}
    \caption{Light curves of event OGLE-2015-BLG-0961 as seen by \emph{Spitzer} and from the ground. All ground-based data sets are shown here. The densely covered ground-based light curve shows no deviation from a point-lens event, which puts an upper limit on the planet-to-star mass ratio $q\lesssim3\times10^{-4}$. The deviation in \emph{Spitzer} light curve would require $q\gtrsim 2\times10^{-3}$. Therefore, the deviation in \emph{Spitzer} data could only be caused by systematics.
    \label{fig:0961}}
\end{figure}

\subsection{Planet Sensitivities} \label{sec:sensitivity}
We apply the planet sensitivity code developed by \citet{Zhu:2015b}. The method was first proposed by \citet{Rhie:2000} 
\footnote{See also the other approach by \citet{Gaudi:2000}.}
and further developed by \citet{Yee:2015b} and \citet{Zhu:2015b} to incorporate space-based observations. Below we provide brief descriptions of the methods and the code, and interested readers can find more details in \citet{Yee:2015b} and \citet{Zhu:2015b}.

The calculation of planet sensitivity requires a certain value for $\rho$, which is the angular source size $\theta_\star$ normalized to $\theta_\e$, $\rho\equiv \theta_\star/\theta_\e$. The angular source size $\theta_\star$ is estimated following the standard procedure, i.e., by comparing the positions of source star and the red clump centroid on the color-magnitude diagram \citep{Yoo:2004}. The determination of $\theta_\e$ follows the prescription given by \citet{Yee:2015b}: for a given solution, we derive the transverse velocity $\tilde{v}_\hel$ using Equation~(\ref{eqn:vvec}), and choose $\mu_\rel=7~\masyr$ if $\tilde{v}_\hel$ favors a disk lens and $\mu_\rel=4~\masyr$ if $\tilde{v}_\hel$ favors a bulge lens; then $\theta_\e=\mu_\rel t_\e$.

We first compute the planet sensitivity $S$ as a function of planet-to-star mass ratio $q$ and the planet/star separation $s$ normalized to the angular Einstein radius $\theta_\e$. Twenty $q$ values are chosen uniformly in logarithmic scale between $10^{-5}$ and $0.04$, which correspond to a mass range from $1~M_\oplus$ to $13~M_{\rm J}$ for a $0.3~M_\odot$ host. Twenty $s$ values are chosen also uniformly in logarithmic scale between $0.3$ to $3$. Our choice of the ``lensing zone'' covers the region where microlensing is sensitive for the nearly all events. For each set of $(q,s)$, we generate 100 planetary light curves that have other parameters the same except for $\alpha$, which is the angle between the source trajectory and the lens binary axis. For each simulated light curve, we then find the best-fit single-lens model using the downhill simplex algorithm, the goodness of which is quantified by $\chi^2_{\rm SL}$. For events that were subjectively chosen and never met the objective criteria, we additionally find the deviation from the single-lens model in the ground-based data that were \emph{released}
\footnote{All KMTNet data were released after the end of the season.}
before the subjective chosen date $t_{\rm sub}$. If this deviation is significant ($\chi^2>10$, \citealt{Yee:2015b}), we consider the injected planet as having been noticeable and thus reject this $\alpha$, regardless of how significant $\chi^2_{\rm SL}$ is. Otherwise, for these events and events that met objective criteria, we pass the simulated events to the anomaly detection filter. The sensitivity $S(q,s)$ is the fraction of $\alpha$ values for which the injected planets are detectable. 

We adopt the following detection thresholds, which are more realistic than that used in \citet{Zhu:2015b} and have been used in \citet{Poleski:2016}: C1. $\chi_{\rm SL}^2>300$ and at least three consecutive data points from the same observatory show $>3\sigma$ deviations; or C2. $\chi^2_{\rm SL}>500$. C1 aims for capturing sharp planetary anomalies, and C2 is supplementary to C1 for recognizing the long-term weak distortions.

In principle, the planet sensitivities could be substantially different for the $\pi_{\e,\pm\pm}$ solutions compared to the $\pi_{\e,\pm\mp}$ solutions, because source trajectories as seen by Earth and \emph{Spitzer} pass by the lens on the projected plane from the same side for the former, but opposite sides for the latter \citep{Zhu:2015b}. However, for the data sets under consideration in the present paper, which typically have several dozen observations per day from the ground and only one or a few per day from space, almost all the sensitivity comes from the ground observations. Hence, the sensitivities of the four degenerate solutions are almost identical. See Figure~6 of \citet{Poleski:2016} for an example. The small differences between four solutions arise from the different values of $\rho$ used in the computation, because $\rho=\theta_\star/(\mu_\rel t_\e)$ and the choice of $\mu_\rel$ relies on the magnitude of $\pi_\e$.

Current experiments are very far from having the ability to separately measure distance distributions for the individual $(s,q)$. Hence, we also define the sensitivity to a given planet-to-star mass ratio $q$
\begin{equation}
    S(q) = \int S(q,s) \dif\log{s}\ .
\end{equation}
This bears the assumption that the distribution of $s$ is flat in logarithmic scale, which is reasonable according to recent studies \citep[e.g.,][]{Fressin:2013,DongZhu:2013,Petigura:2013,Burke:2015,Clanton:2016}.

\section{Results} \label{sec:results}

\subsection{Light Curves \& Systematics} \label{sec:systematics}
We present the ground-based and space-based light curves of each event in our sample in Figures~\ref{fig:lcs-1}, \ref{fig:lcs-2}, and \ref{fig:lcs-3}. All data sets except OGLE have been re-scaled to the OGLE $I$ magnitude based on the best-fit model
\begin{equation} \label{eqn:rescaling}
    \tilde{F}^{j} = \frac{F_{\rm s}^{\rm OGLE}}{F_{\rm s}^{j}}(F^{j}-F_{\rm b}^{j}) + F_{\rm b}^{\rm OGLE}\ .
\end{equation}
We suppress KMTNet data sets in these plots, and only show OGLE data for clarity. The reader can find an example event that demonstrates the much denser coverage of KMTNet in Figure~\ref{fig:0961}.

The ground-based data of all 50 events in our sample can be well fitted by a single-lens model. However, the \emph{Spitzer} data of several of them show deviations from this simple description. Some of these deviations are prominent, such as in OGLE-2015-BLG-0081, 0461, 0703, 0961 and 1189. However, we believe that these are due to unknown systematics in the \emph{Spitzer} data rather than indications of companions to the lens. Below we provide two examples to demonstrate this point. \citet{Poleski:2016} noticed a strong deviation in the \emph{Spitzer} data of OGLE-2015-BLG-0448. Although they found that a lens companion with $q=1.7\times10^{-4}$ could improve the single-lens model by $\Delta \chi^2=128$, they showed that even the best-fit binary-lens model could not remove all the deviations in the \emph{Spitzer} data. Therefore, the trend in \emph{Spitzer} data was likely caused by systematics rather than physical signal from additional lens object. This is especially true for OGLE-2015-BLG-0961. As shown in Figure~\ref{fig:0961}, the ground-based data can be well fitted by a single-lens model with extremely high magnification ($u_{0,\oplus}\le0.005$ at 1-$\sigma$ level), which excludes any lens companions with $q\gtrsim 3\times10^{-4}$ if close to the Einstein ring (see Figure~\ref{fig:sens-1}). The \emph{Spitzer} data show a long term deviation centered at the time when the event peaked from the ground. 
{This long term deviation, if attributed to a companion to the lens, could only be caused by the planetary caustic. With $u_{0,spitz}=0.1$ and the position of planetary caustic at $|s-1/s|$, the separation between the hypothetical lens companion and the primary lens should be $\log{s}=\pm0.02$. Combining the duration of the deviation (10~days out of $t_\e=60$~days) and the width of planetary caustic \citep{Han:2006}, we can put limit on the companion mass ratio $q\gtrsim 2\times10^{-3}$.}
There do not exist any $q$ values that could explain the non-detection in the ground-based data and the significant trend in \emph{Spitzer} data. Therefore, the trend in \emph{Spitzer} data is likely due to systematics in the \emph{Spitzer} data reduction.
\footnote{In principle, the trend in \emph{Spitzer} data can also be caused by binary sources. However, this scenario requires a secondary source that is nearly as faint as the primary source, but redder by 2.5 mag in $I-[3.6\microm]$ or 1.6 mag in $V-I$. Such stars are extremely rare. Therefore, it is very unlikely that the trend is caused by binary sources.}

The systematics in \emph{Spitzer} data can potentially affect the parallax measurements. However, it has been demonstrated that the influence is small in several published events. For example, \citet{Poleski:2016} showed that the parallax parameters with and without the systematic trend were almost identical. The agreement between orbital parallax and satellite parallax also indicates that the effect of systematics is less likely an issue \citep[e.g.,][]{Udalski:2015a,Han:2017}.

\begin{figure*}
    \epsscale{1.05}
    \plotone{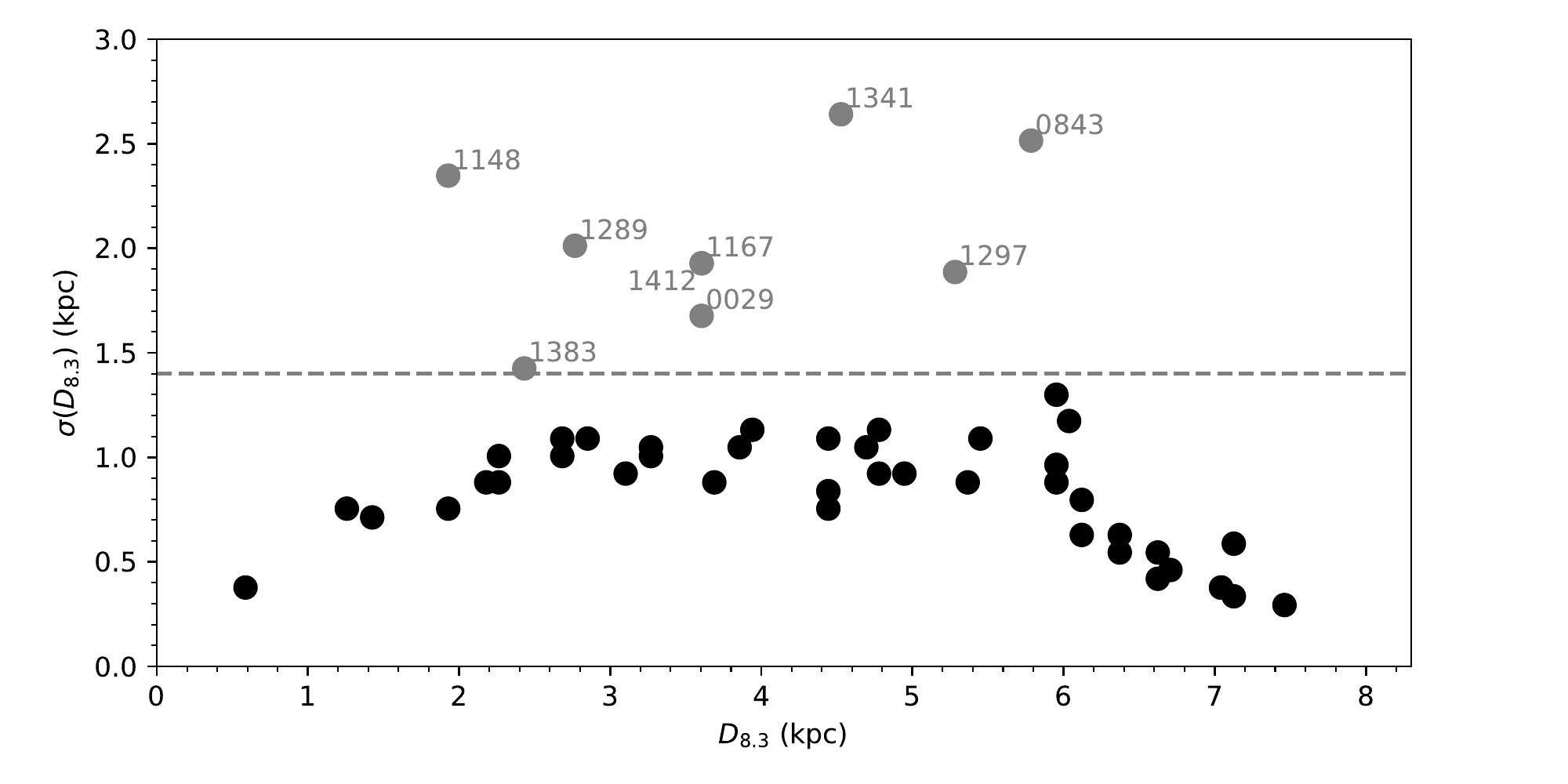}
    \caption{The median and the 1-$\sigma$ uncertainty of the lens distance parameter $\Dparm$ (Equation~\ref{eqn:ddef}) of all 50 events in our raw sample. We exclude events with $\sigma(\Dparm)>1.4~$kpc from the final statistical sample. This criterion was adopted based on examination of the distributions of $\Dparm$, which are shown in Figure~\ref{fig:lens-dists}. The OGLE numbers of all excluded events are labeled.
    \label{fig:measure-pie}}
\end{figure*}

\begin{figure*}
    \epsscale{1.05}
    \plotone{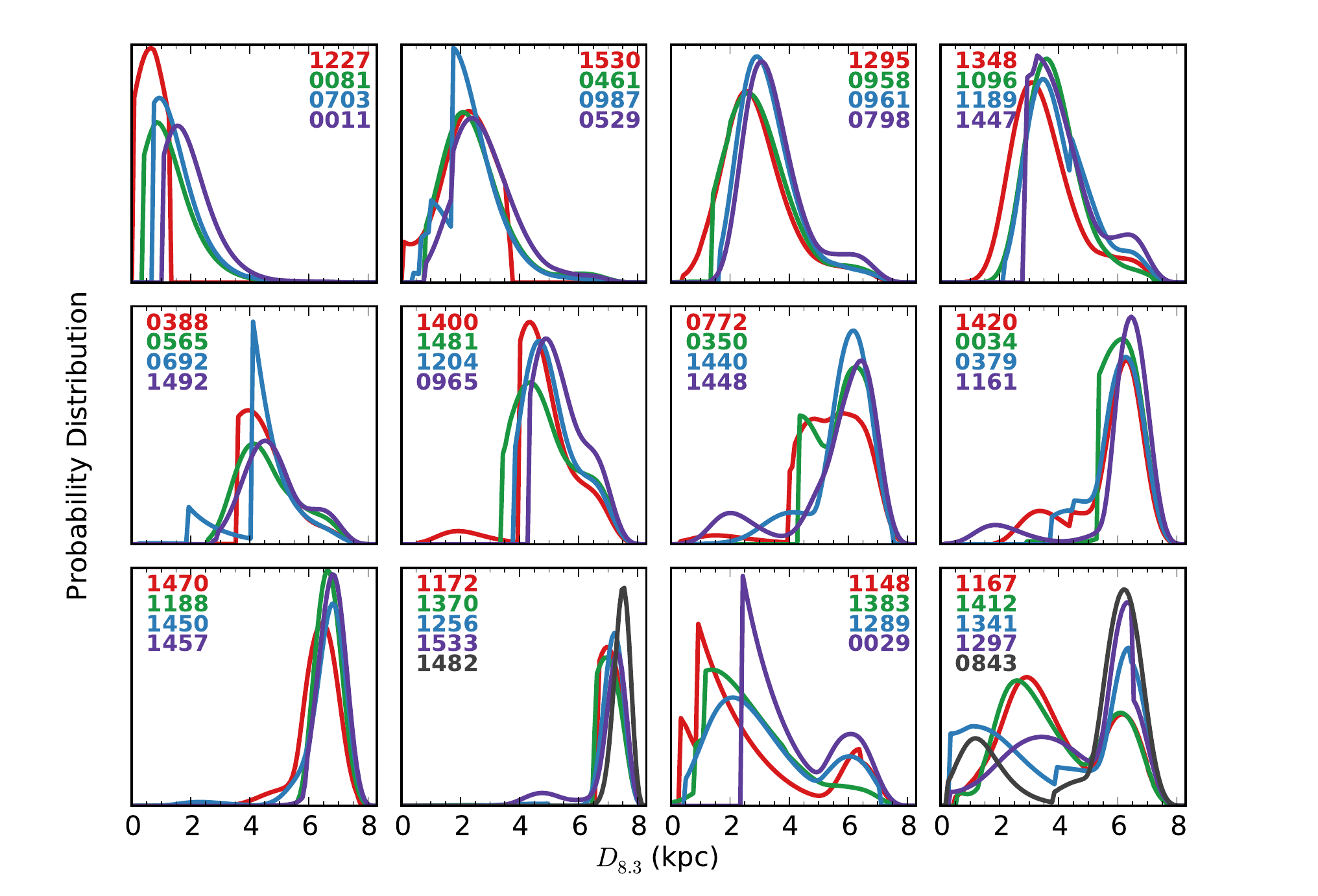}
    \caption{The distributions of lens distance parameter $\Dparm$ for all 50 events in our raw sample. Events in the last two panels (bottom right) are excluded from the final sample because of their broad $\Dparm$ distribution. Events included in the final sample, as well as events excluded from the final sample, are shown in the order of increasing median $\Dparm$.
    \label{fig:lens-dists}}
\end{figure*}

\begin{figure*}
    \epsscale{1.05}
    \plotone{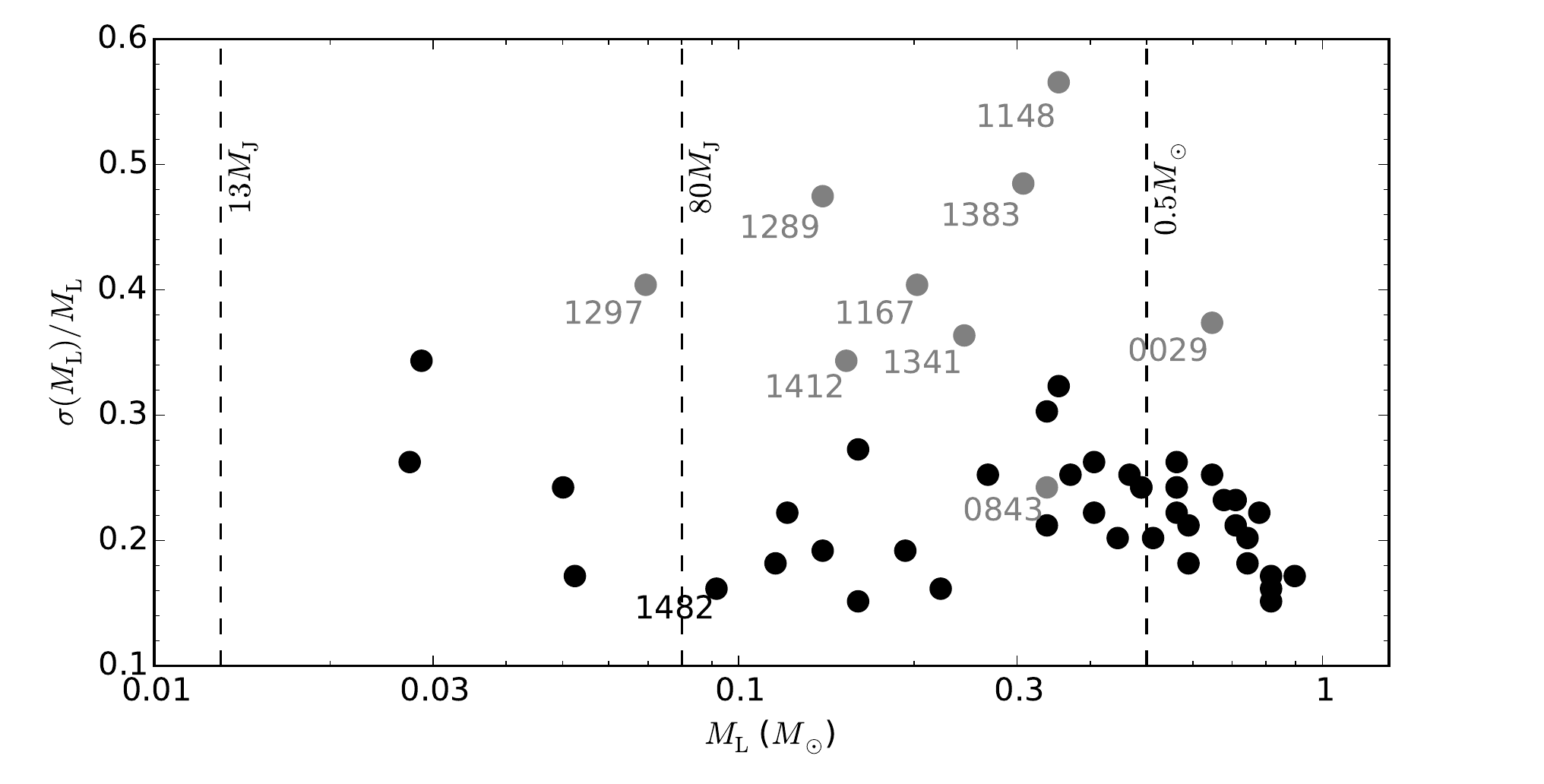}
    \caption{The median and the fractional uncertainty of the lens mass $M_\l$ of all 50 events in our raw sample. Events that are excluded from the final sample based on $\sigma(\Dparm)$ criterion are shown in gray and have their OGLE numbers labeled aside. The vertical dashed lines indicate three characteristic masses, $13~M_{\rm J}$, $0.08~M_\odot$, and $0.5~M_\odot$, respectively. Event OGLE-2015-BLG-1482 has direct mass measurement from the finite-source effect, $M_\l=0.10\pm0.02~M_\odot$ or $0.06\pm0.01~M_\odot$ \citep{Chung:2017}. Our Bayesian estimate of the mass agrees with the direct measurement pretty well {($\le2~\sigma$)}.
    \label{fig:mass-scatter}}
\end{figure*}

\begin{figure*}
    \epsscale{1.05}
    \plotone{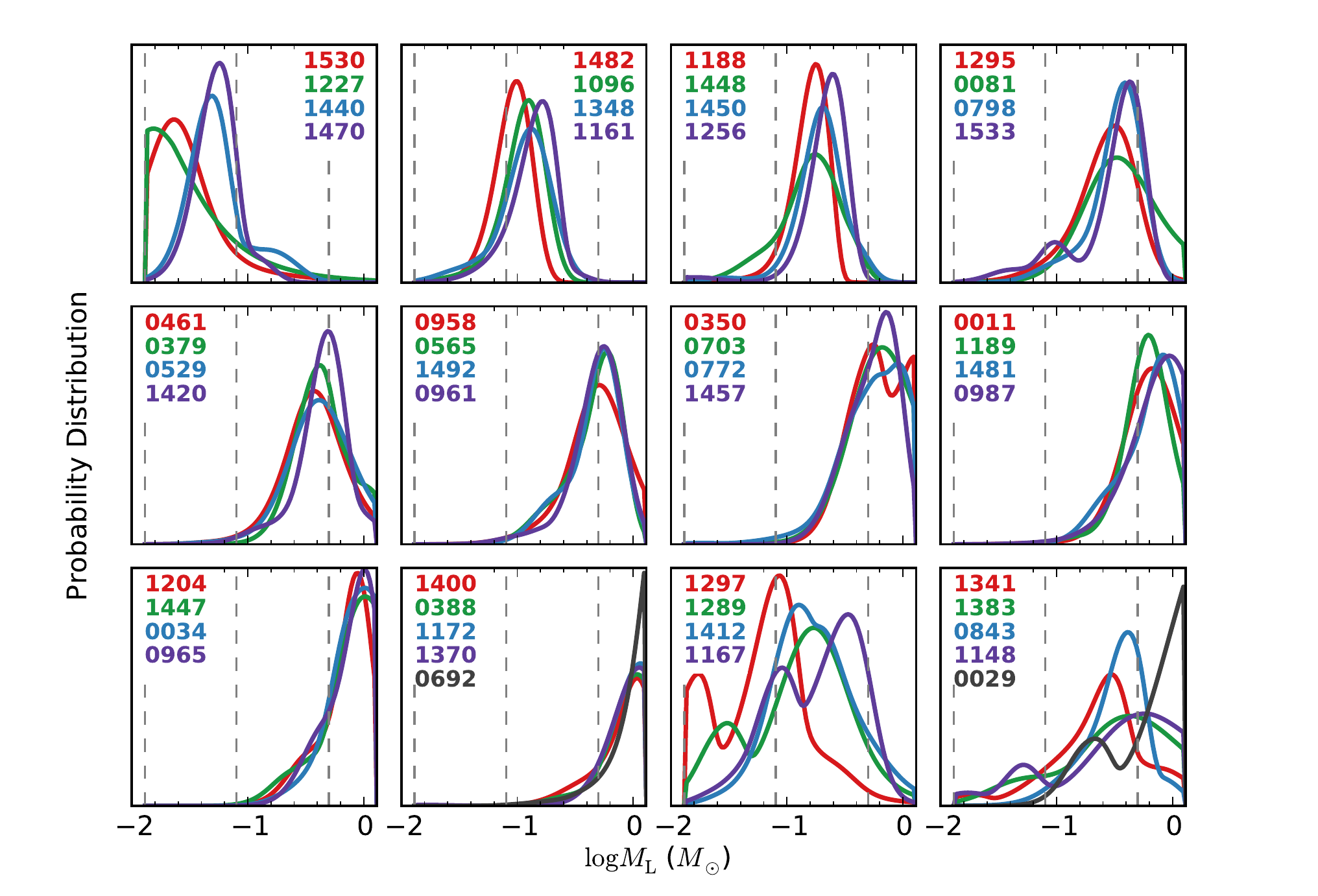}
    \caption{The distributions of lens mass $M_\l$ for all 50 events in our raw sample. Events in the last two panels (bottom right) are excluded from the final sample because of their broad $\Dparm$ distribution. Events included in the final sample, as well as events excluded from the final sample, are shown in the order of increasing $M_\l$ median. The vertical dashed lines indicate three characteristic masses, $13~M_{\rm J}$, $0.08~M_\odot$, and $0.5~M_\odot$, respectively.
    \label{fig:lens-mass}}
\end{figure*}

\begin{figure*}
    \epsscale{1.2}
    \plotone{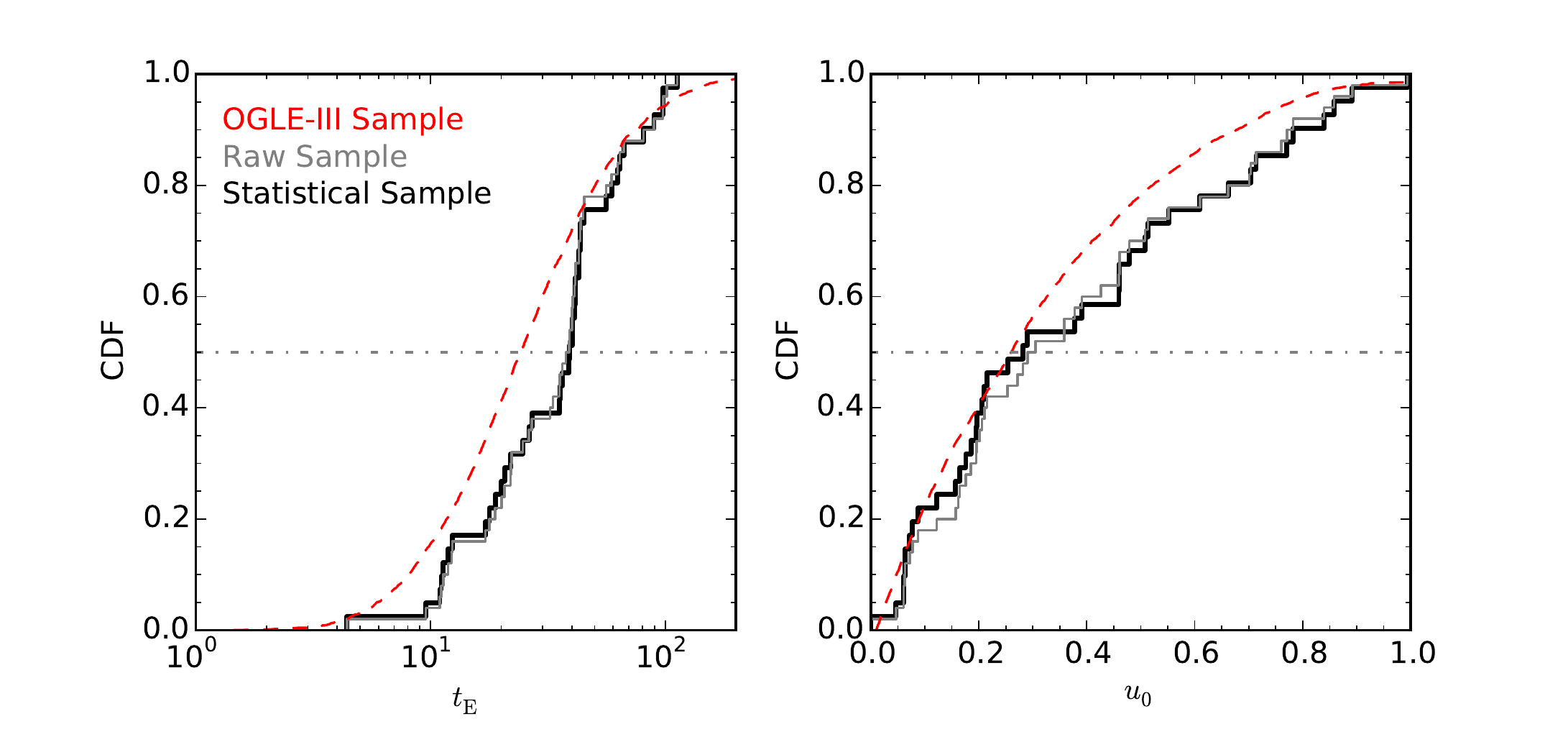}
    \caption{Cumulative distributions of timescale $t_\e$ and impact parameter $u_0$ for events in our sample (black solid curves) and in the OGLE-III sample (red dashed curves) from \citet{Wyrzykowski:2015}. Events with $u_0<0.01$ in the OGLE-III sample have been excluded because of their unreliable parameters \citep{Gould:2010}. For each event in our sample, the values are chosen from the solution that has the lowest $\chi^2$, although the differences between different solutions are small. The gray horizontal lines indicate the median level.
    \label{fig:teu0}}
\end{figure*}

\begin{figure*}
    \epsscale{1.2}
    \plotone{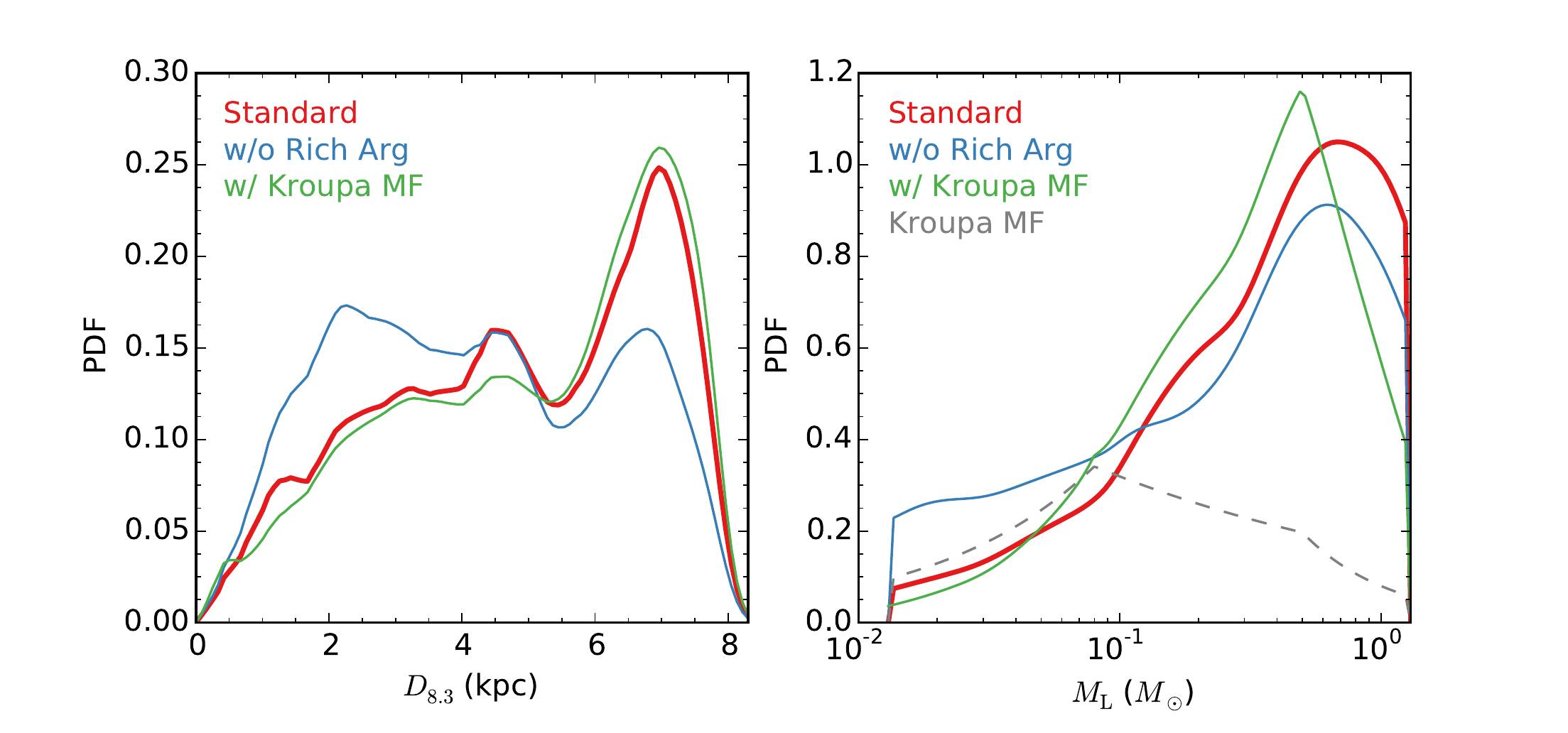}
    \caption{The differential probability distribution functions (PDF) of lens distance parameter $\Dparm$ (left panel) and lens mass $M_\l$ for the 41 events in our sample. We choose the results with flat MF (in $log{M_\l}$) and Rich argument as ``standard'', but also consider cases in which the Rich argument is removed (labeled ``w/o Rich Arg'') and the MF is replaced with the Kroupa MF (labeled ``w/ Kroupa MF''), respectively. Note, in particular, that changing the mass function has almost no effect on the inferred distances. In the right panel we also illustrate the Kroupa MF (Equation~\ref{eqn:kroupa}), employing an arbitrary normalization for this purpose.
    \label{fig:bayes}}
\end{figure*}

\subsection{Event Parameters \& Lens Distributions} \label{sec:measure-pie}

We provide in Table~\ref{tab:parameters} the best-fit parameters as well as associated uncertainties for solutions with $\Delta \chi^2\le100$ of all 50 events. Here $\Delta \chi^2$ is the difference between a given solution and the best solution for that event. With these, and following the method in Section~\ref{sec:bayes}, we derive the lens distance parameter $\Dparm$ and lens mass $M_\l$ distributions for every event in our raw sample.

Based on all event parameters and the subsequent lens distributions, we can now select events for our final statistical sample. The guideline is that only events with ``detected parallax'' can be included for the study of the Galactic distribution of planets, as \citet{Yee:2015b} pointed out. At first sight, the above guideline seems to suggest a criterion on the measurement uncertainty of $\bdv{\pi_\e}$. However, such an approach would be problematic, in particular because the uncertainty of $\bdv{\pi_\e}$ is determined for individual solution, but decisions have to be made for individual events, which generally have more than one solutions. As shown in Table~\ref{tab:parameters}, the $(\pm,\mp)$ solutions are in general better constrained than the $(\pm,\pm)$ solutions, so even though they are statistically disfavored by the Rich argument, they are more likely to survive if a cut on the detection significance of $\pi_\e$ is applied.
Although it is possible to design a criterion for choosing events that balances the two opposite factors, a better approach is to choose events based on the distance parameter $\Dparm$ and its associated uncertainty $\sigma(\Dparm)$. This is because only events with well determined distances contribute to the measurement of the Galactic distribution of planets.

We show in Figure~\ref{fig:measure-pie} the median value and the 1-$\sigma$ uncertainty of the lens distance parameter $\Dparm$ derived for each event in our raw sample. Here the 1-$\sigma$ uncertainty is the half-width of a 68\% confidence interval centered on the median $\Dparm$. By visually inspecting the $\Dparm$ distributions of all 50 events, which are shown in Figure~\ref{fig:lens-dists}, we decide to use $\sigma(\Dparm)\le 1.4~$kpc as the criterion for claiming a parallax detection and thus for any event to be included in the final sample. 
We end up with 41 events in the final sample. The 9 events that are excluded all have broad distributions of $\Dparm$, even though some of them have very good measurements of $\pi_\e$ (e.g., OGLE-2015-BLG-0029, 0843, 1167). The broad distribution of $\Dparm$ arises from the atypical magnitude and direction of $\bdv{\tilde{v}_\hel}$. When combined with the Galactic model, the former favors near- to mid-disk lenses while the latter favors bulge lenses.

The derived lens masses and the fractional uncertainties are shown in Figure~\ref{fig:mass-scatter}. As expected, events that do not show compact $\Dparm$ distributions do not have well constrained mass estimates, either. For events in our final sample, the typical uncertainty of the lens mass estimate is 20\%, regardless of it is substellar or not. In particular, we note that the lens mass estimate of OGLE-2015-BLG-1482 agrees reasonably well {($\le2~\sigma$)} with the direct mass measurement from the finite-source effect \citep{Chung:2017}, as a demonstration that the mass estimate method employed here is valid. The derived lens mass distributions of all 50 events are presented in Figure~\ref{fig:lens-mass}.

\begin{figure*}
    \centering
    \epsscale{1.1}
    \plotone{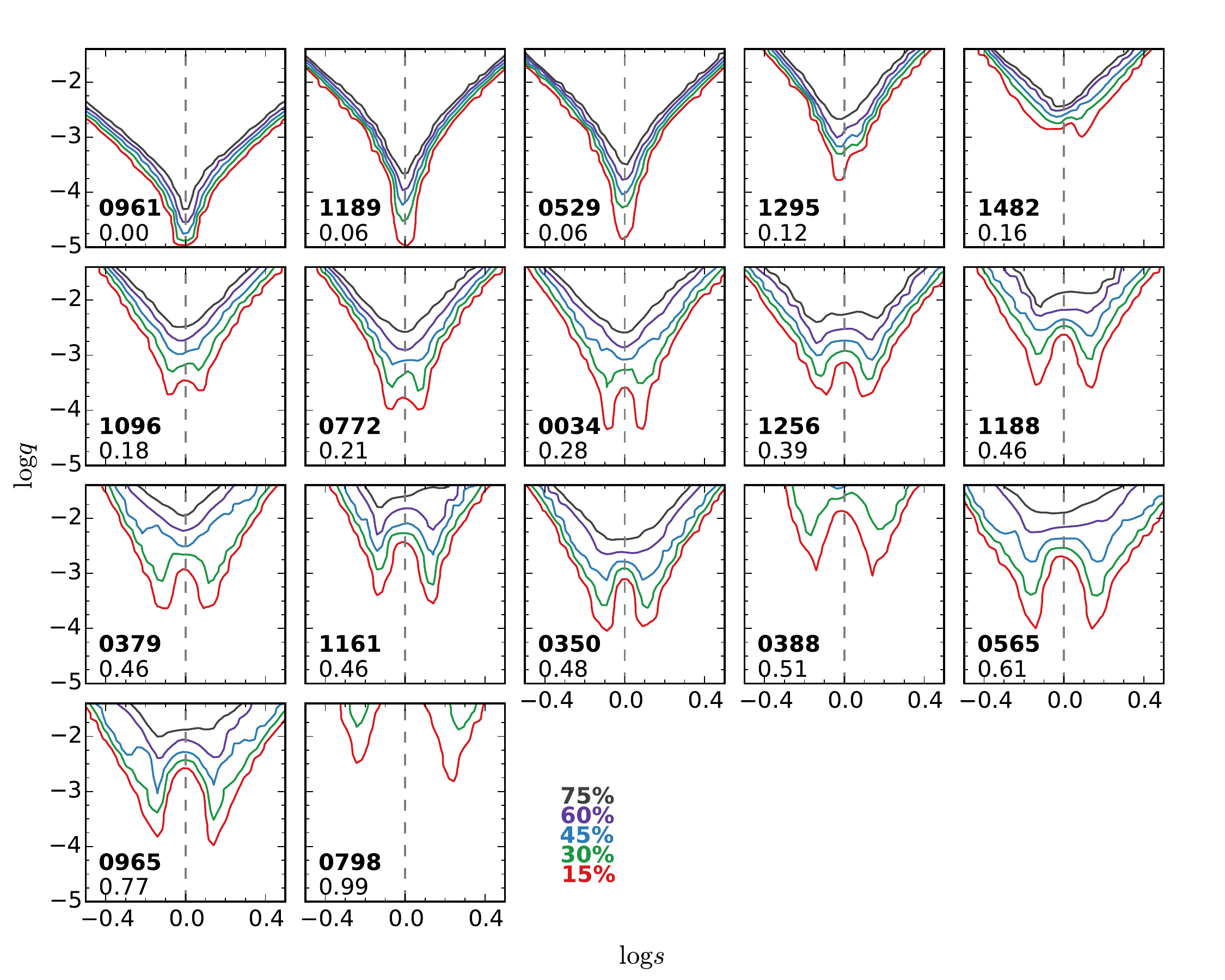}
    \caption{Planet sensitivity curves of the 17 objectively selected events, sorted by the impact parameters. The OGLE number and the impact parameter are provided at the lower left corner in each plot. The colors represent the curves with different sensitivities in $S(q,s)$. For simplicity, we only show the sensitivity curves for the $(+,+)$ solution, regardless of how many solutions we calculated. The difference between sensitivity curves of different solutions is small.
    \label{fig:sens-1}}
\end{figure*}

\begin{figure*}
    \centering
    \epsscale{1.1}
    \plotone{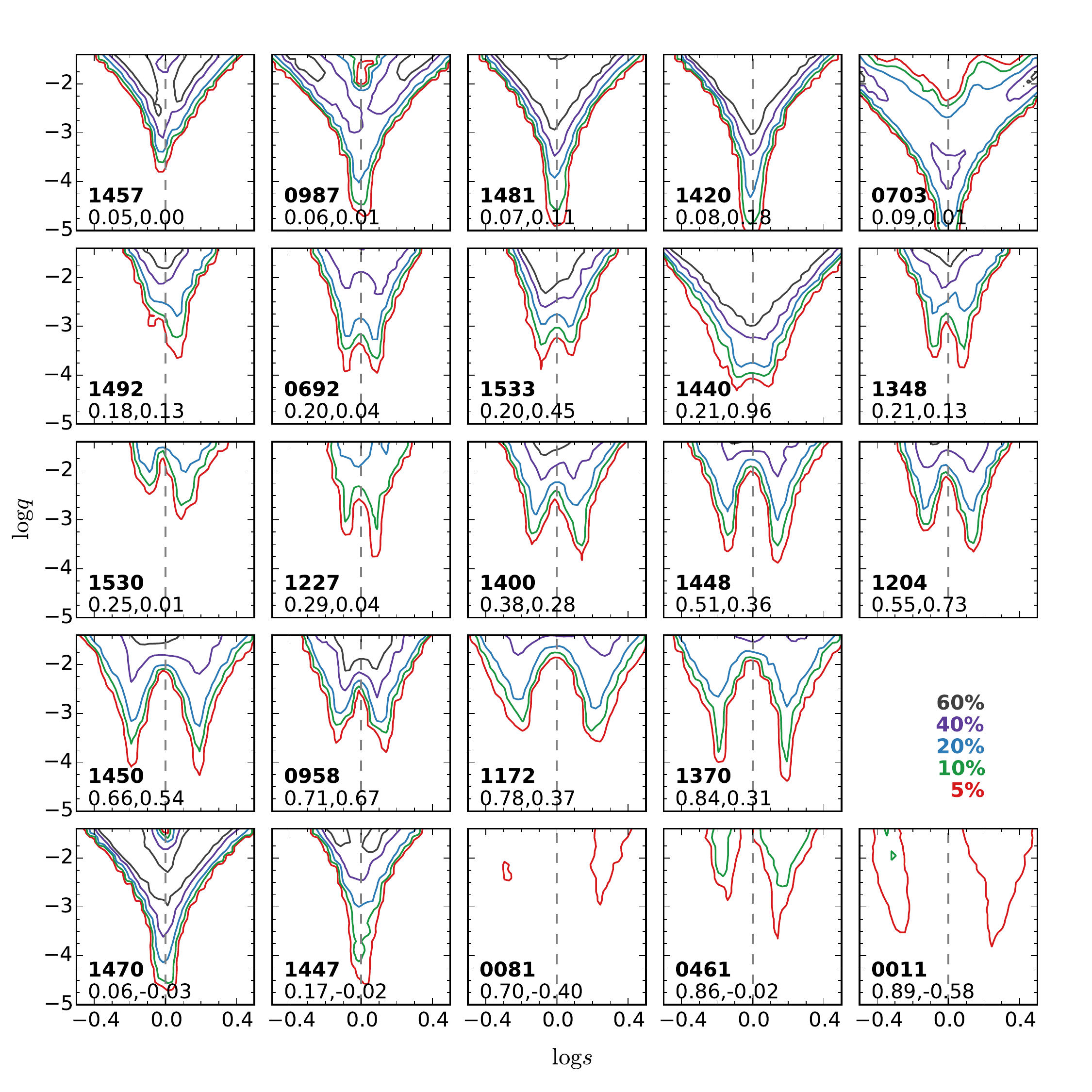}
    \caption{Planet sensitivity curves of the 24 subjectively selected events. These events are grouped into two categories: events selected before the peak as seen from the ground (top four rows), and events selected after the peak as seen from the ground (bottom row). In each category, events are shown in the order of increasing impact parameter. In each panel, we indicate the OGLE number (in bold), impact parameter $u_0$, and the subjective selection relative to the peak, $(t_0-t_{\rm sub})/t_\e$, at the lower left corner.
    \label{fig:sens-2}}
\end{figure*}

\begin{figure*}
    \centering
    \epsscale{1.}
    \plotone{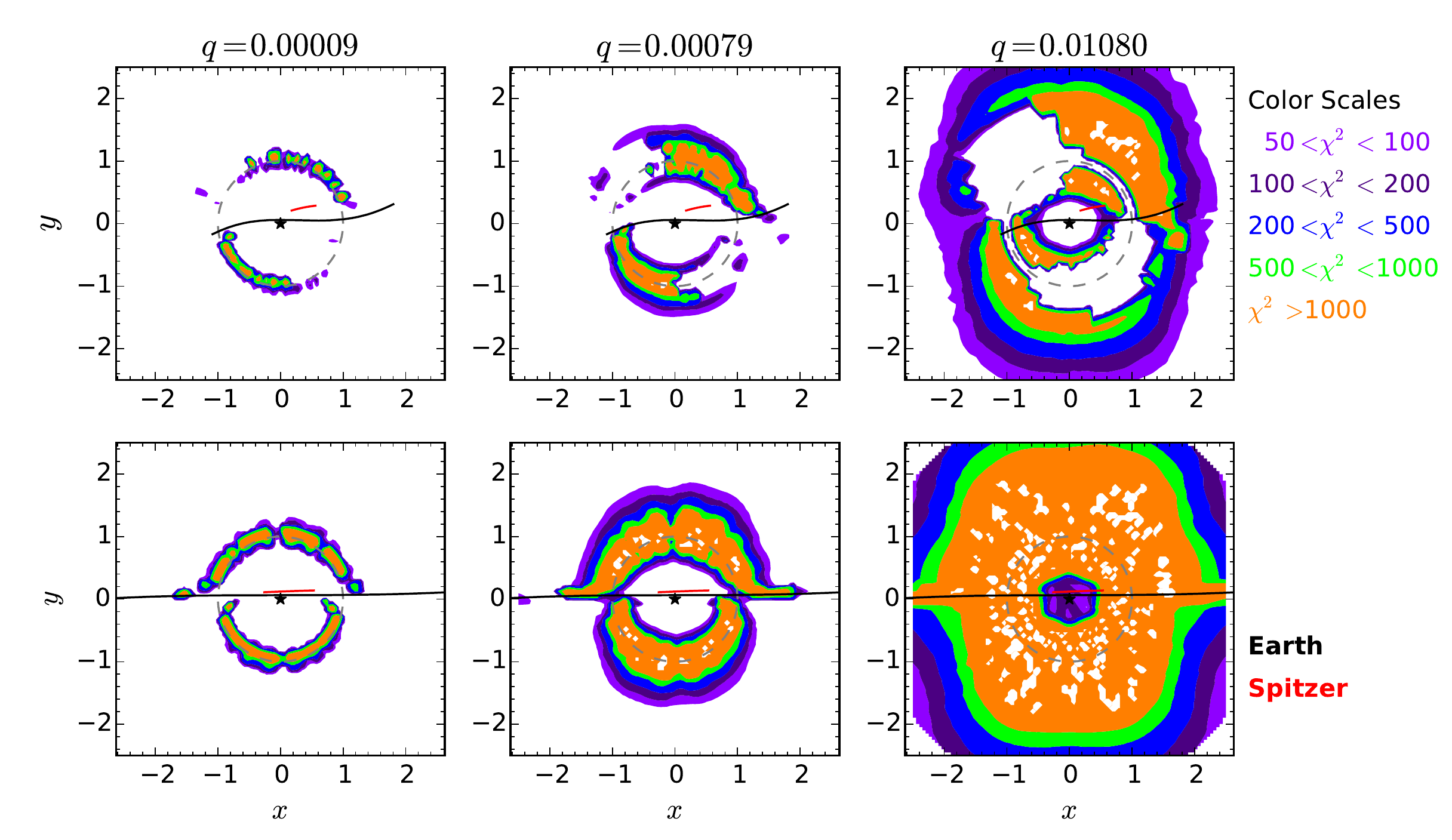}
    \caption{The detectabilities (i.e., the $\chi^2$ deviation from a single-lens light curve) of three $q$ values at different positions $(x,y)$ for two events OGLE-2015-BLG-0987 (upper panels) and OGLE-2015-BLG-1189 (lower panels). These two events have similar impact parameters $u_{0,\oplus}\approx0.06$, but their sensitivities to planets are quite different due to their different selection statuses. See the text for more explanations. In each panel, the red and black curves indicate the source trajectories seen by \emph{Spitzer} and from the ground, respectively, and the gray dashed curve indicates the position of Einstein ring.
    \label{fig:chi2map-demo}}
\end{figure*}

We show in Figure~\ref{fig:teu0} the cumulative distributions of event timescales $t_\e$ and impact parameters $u_{0,\oplus}$ as seen from the ground, and compare them with those in the OGLE-III microlensing event catalog \citep{Wyrzykowski:2015}, which can be considered to be complete and uniform for our purpose. Because different solutions have only slightly different $t_\e$ and $u_{0,\oplus}$, we simply take values of the solution with lowest $\chi^2$. For the timescale $t_\e$ distribution, we notice that events in our final sample are more concentrated within 10--100 days than events in the OGLE-III catalog. The lower limit comes into play because there is a 3--9 day lag between events being selected and events getting observed by \emph{Spitzer} (see Figure~1 of \citealt{Udalski:2015a}). The lack of extremely long timescale ($t_\e \gtrsim100$~days) events comes as a consequence of our event selection criteria, because a substantial brightness change ($\gtrsim$0.3 mag) during the $\sim$40-day \emph{Spitzer} bulge window is required in order to detect the parallax effect \citep{Yee:2015b}. 
Although the events in our sample (and subsequent larger samples) have a biased $t_\e$ distribution, this bias applies to both events with and without planet detections in the same way. Therefore, it will not affect the statistical studies of the Galactic distribution of planets.
The \emph{Spitzer} sample $u_0$ distribution shows similar overall morphology to that of the OGLE-III catalog, but is more uniform, which indicates that it shows less magnification bias. This again reflects that the fact OGLE-III detections are possible based on a few days of relatively magnified sources, whereas \emph{Spitzer} selections are delayed by 3--9 days.

We show in Figure~\ref{fig:bayes} the distributions of lens distance parameter $\Dparm$ and lens mass $M_\l$, which are averaged over all 41 events in the final sample.
We consider the influences of the Rich argument and a different choice of stellar mass function (e.g., Kroupa MF). As expected (see Section~\ref{sec:solutions}, also \citealt{SCN:2015a}), the lens distance distribution is biased toward more nearby and therefore lower-mass lenses, if the Rich argument is not taken into account. The different choices of the stellar mass function have marginal effect, especially on the lens distance distribution. Our result demonstrates, for the first time, that the peak of the microlens mass distribution is at $0.5~M_\odot$, and that the majority microlensing events are caused by M-dwarfs.

\subsection{Planet Sensitivities \& Constraints on Planet Distribution Function}

We present in Figures~\ref{fig:sens-1} and \ref{fig:sens-2} the planet sensitivity plots of individual events in our final sample. 
Events are divided according to their final status of \emph{Spitzer} selections, with objectively selected events shown in Figure~\ref{fig:sens-1} and subjectively selected events shown in Figure~\ref{fig:sens-2}.
For all objective events and most subjective events, the sensitivity curves are smooth and triangle-like, with either a single horn (for relatively high magnification events, see also \citealt{Gould:2010}) or double horns (for relatively low magnification events, see also \citealt{Gaudi:2002}). In the remaining subjective events, however, the sensitivity curves show discontinuity especially at large $q$ values. This was caused by the way that the planet sensitivity of subjectively chosen event was computed. As described in detail in \citet{Yee:2015b} and \citet{Zhu:2015b}, and summarized in Section~\ref{sec:sensitivity}, for events that were chosen subjectively and never met the objective selection criteria, all (hypothetical) planet detections must be censored from the statistical sample if they would have betrayed their existence in the data that were released before the subjective selection date $t_\sub$. This has only a marginal effect if $t_\sub$ is well before the event peak $t_{0,\oplus}$, because the bulk of planet sensitivities come from the region near the peak ($|t-t_{0,\oplus}|\lesssim u_{0,\oplus}t_\e$). If $t_\sub$ is close to $t_{0,\oplus}$, then the above procedure could affect the final sensitivity curves significantly. In particular, planets that are more massive and closer to the Einstein ring are more easily excluded in the sensitivity computation; for given combinations of $q$ and $s$, some choices of $\alpha$ are more easily discarded as well. As an example, we show in Figure~\ref{fig:chi2map-demo} the $\chi^2$ maps for three different $q$ values for two events, OGLE-2015-BLG-0987 and OGLE-2015-BLG-1189, which have similar impact parameters $u_{0,\oplus}$ but show very different sensitivity curves.

\begin{figure*}
    \epsscale{1.2}
    \plotone{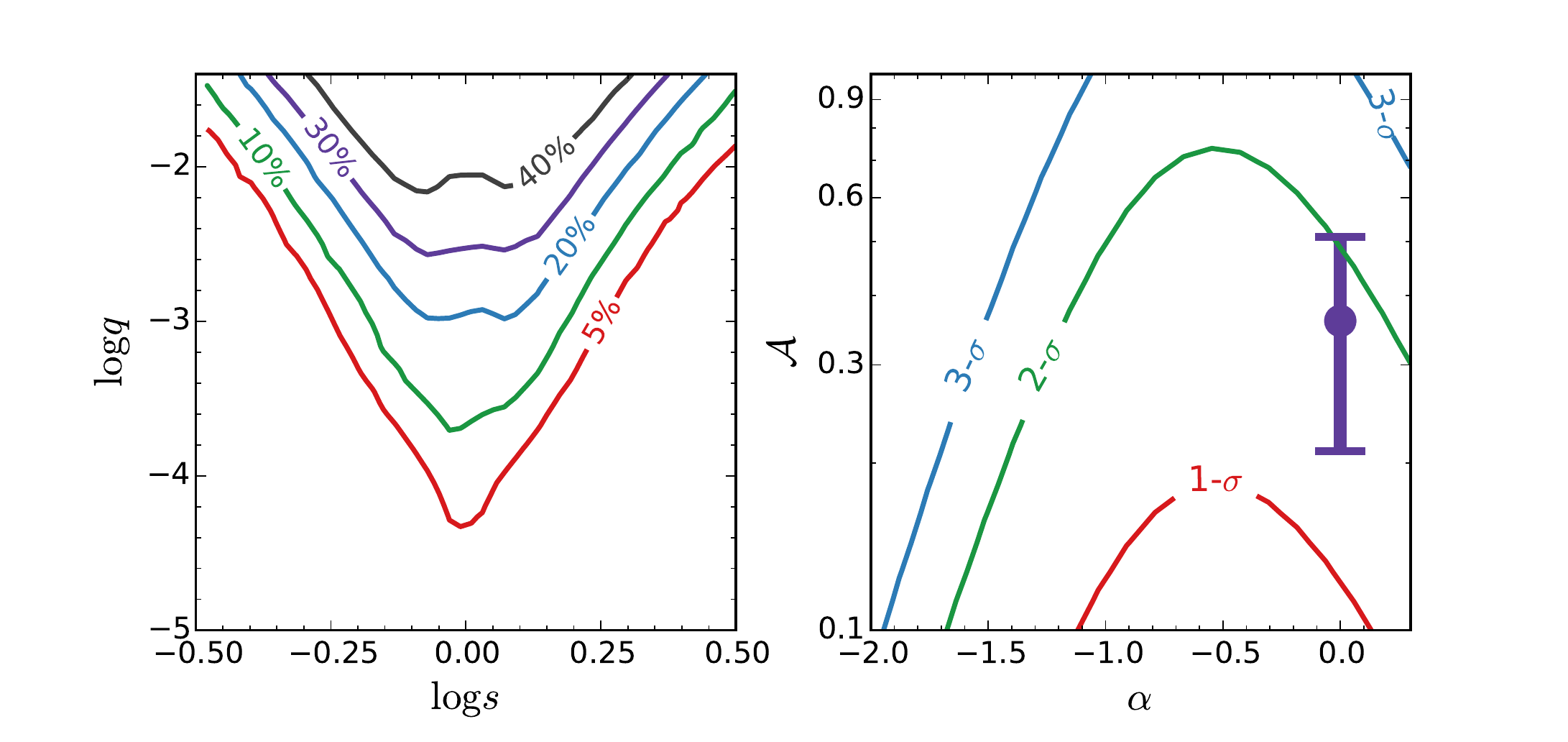}
    \caption{\textit{Left panel}: planet sensitivities averaged over 41 events in our sample. \textit{Right panel}: constraints on the planet distribution function based on our sample. Here $\alpha$ is the slope of the planet mass function, and $\mathcal{A}$ is the normalization factor. The purple point is the measurement by \citet{Gould:2010}, which assumed flat ($\alpha=0$) planet mass function.
    \label{fig:yields}}
\end{figure*}

We provide constraints on the planet distribution function, based on the null detection in our sample. We adopt the following form as the planet distribution function
\begin{equation} \label{eqn:pdf}
    \frac{\dif N}{\dif \log{q}} = \mathcal{A} \left(\frac{q}{q_{\rm ref}}\right)^{\alpha}\ ,
\end{equation}
and choose $q_{\rm ref}=5\times10^{-4}$, which is the typical $q$ value of microlensing planets \citep[e.g.,][]{Gould:2010}. We first show on the left panel of Figure~\ref{fig:yields} the sensitivity curves averaged over the 41 events in the final sample. Assuming Poisson-like noise and that ``planets'' should have $q\le 10^{-2}$ (to be consistent with previous studies, e.g., \citealt{Gould:2010}), we are able to derive the constraints on the slope of the planet mass function $\alpha$ and the normalization factor $\mathcal{A}$ based on the null detection in our sample. The results are shown on the right panel of Figure~\ref{fig:yields}. 
Our constraints are consistent, at 2-$\sigma$ level, with previous statistical studies based on samples of microlensing planets \citep{Gould:2010,Shvartzvald:2016,Suzuki:2016}.
In particular, we find $\mathcal{A}<0.49$ at 95\% confidence level for a flat ($\alpha=0$) planet mass function, which is consistent with the result ($\mathcal{A}=0.36\pm0.15$) from \citet{Gould:2010}. 

\begin{figure*}[h!]
    \epsscale{1.2}
    \plotone{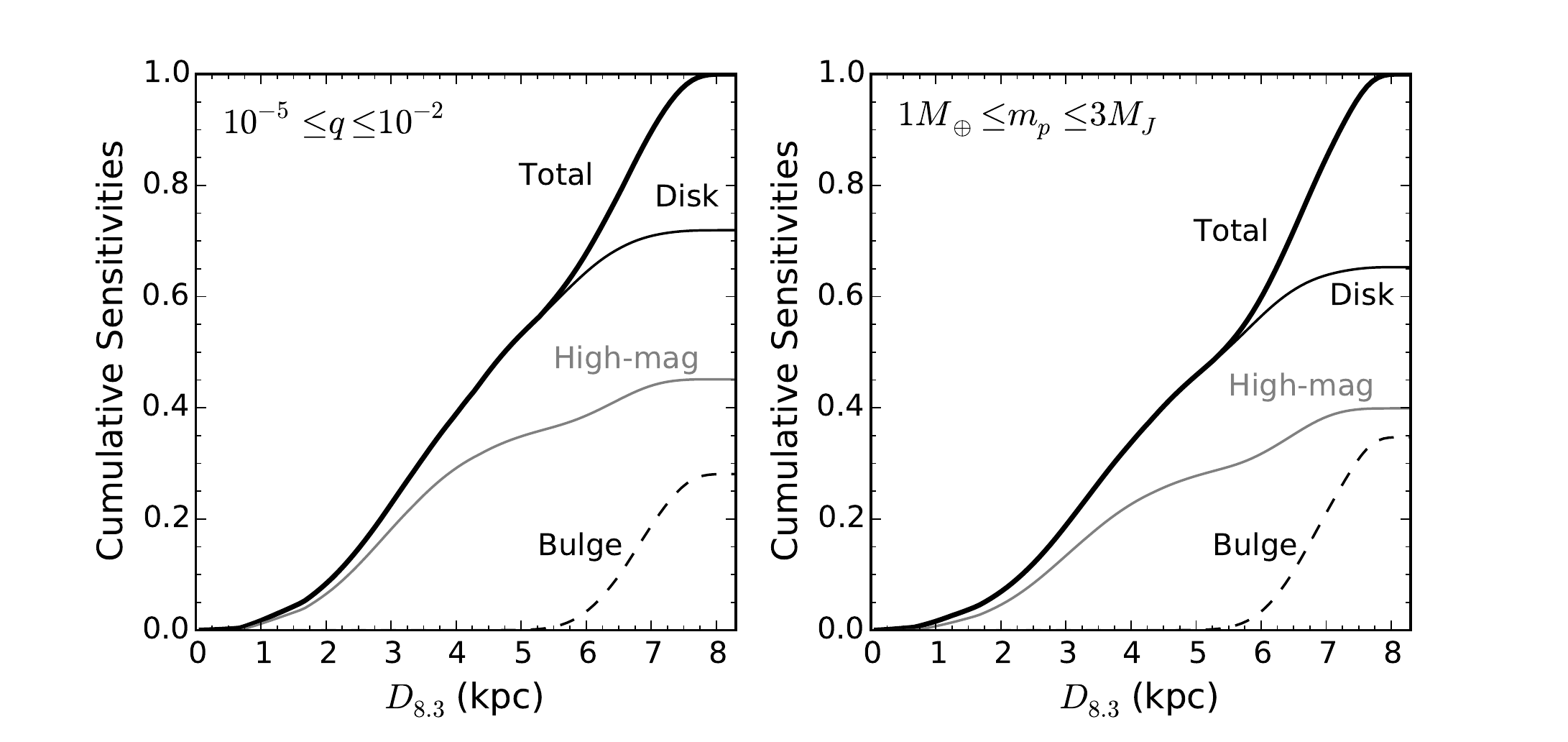}
    \caption{The normalized cumulative planet sensitivities along the lens distance parameter $\Dparm$. Planets are defined by $10^{-5}\le q\le10^{-2}$ in the left panel and by $1~M_\oplus\le m_p\le 3~M_\jup$ in the right panel. Contributions from the disk events, bulge events, and relatively high-magnification ($A_{\rm max}>8$) events are also shown separately.
    \label{fig:cumu-sens}}
\end{figure*}

\subsection{Galactic Distribution of Planets}

We derive the cumulative distribution of planet sensitivities of our sample based on the lens distribution $P(\Dparm)$ and the planet sensitivity $S(q)$. The results are presented here in terms of both the planet-to-star mass ratio $q$
\begin{equation} \label{eqn:cdf-q}
    \small
    \mathcal{C}_q(\Dparm) = \frac{1}{\mathcal{C}_q(\Rgc)} \sum_{i,j} \int_0^{\Dparm} P_i^j(D')\dif D' \int_{q_\min}^{q_\max} S_i^j(q) \dif\log{q}\ ,
\end{equation}
and the planet mass $m_\p$
\begin{equation} \label{eqn:cdf-m}
    \small
    \mathcal{C}_m(\Dparm) = \frac{1}{\mathcal{C}_m(\Rgc)} \sum_{i,j} \int_0^{\Dparm} P_i^j(D')\dif D' \int_{q_\min(D')}^{q_\max(D')} S_i^j(q) \dif\log{q}\ .
\end{equation}
Here $P_i^j(\Dparm)$ and $S_i^j(q)$ are the lens distance distribution and the planet sensitivity for solution $i$ of event $j$. In Equation~(\ref{eqn:cdf-q}), we choose $q_\min=10^{-5}$ and $q_\max=10^{-2}$. In Equation~(\ref{eqn:cdf-m}), we solve for the boundaries on $q$ for individual $\Dparm$ values that lead to the planet mass ranging from $1~M_\oplus$ to $3~M_\jup$. These two distributions are normalized so that $\mathcal{C}(\Rgc)=1$. The results are shown as the solid black curves in Figure~\ref{fig:cumu-sens}.

In terms of the Galactic distribution of planets, we derive the ratio of planets in the bulge to planets in the disk {(to which our survey is sensitive)}
\begin{equation}
    \eta_{\b2d} \equiv \frac{\int_0^{\Rgc} \mathcal{C}'(\Dparm) f_B(\Dparm) \dif \Dparm}{\int_0^{\Rgc} \mathcal{C}'(\Dparm) [1-f_B(\Dparm)] \dif \Dparm}\ .
\end{equation}
Here $\mathcal{C}'(\Dparm)$ is the derivative of $\mathcal{C}(\Dparm)$, and $f_B(\Dparm)$ is the contribution of bulge events (lens in the bulge) to all events at given $\Dparm$, which is given by
\begin{equation} \label{eqn:bulge-frac}
    f_B(\Dparm) = \frac{\int_{D_{\s,\min}}^{D_{\s,\max}} n_{\rm B}(D_\l)/n_\star(D_\l) n_\star(D_\s) D_\s^{2-\gamma} D_\l^2 \dif D_\s}{\int_{D_{\s,min}}^{D_{\s,\max}} n_\star(D_\s) D_\s^{2-\gamma} D_\l^2 \dif D_\s}\ .
\end{equation}
Here $D_\l$ is derived for given $\Dparm$ and $D_\s$. As illustrated in Figure~\ref{fig:cumu-sens}, for the current sample we find that $\eta_{\b2d}=28\%$ if ``planet'' is defined by mass ratio $q$ in the range $10^{-5}$ to $10^{-2}$, and that $\eta_{\b2d}=35\%$ if ``planet'' is defined by mass in the range $1~M_\oplus$ to $3~M_\jup$. Assuming the planet formation is no different between the bulge and the disk, these suggest that $\sim$1/3 of all planet detections in our experiment should come from bulge events. In other words, any deviation from the above value would indicate that the bulge planet population is different from the disk planet population.

We also investigate the influence of impact parameters on the two cumulative distributions $\mathcal{C}_q(\Dparm)$ and $\mathcal{C}_m(\Dparm)$, for the purpose of better planning future similar experiments. We find that the sample of events with maximum magnifications $A_\max>8$, which account for 24\% of all events in our sample, contributes 40\% to 45\% of all planet sensitivities.

\section{Discussion} \label{sec:discussion}

We present the planet sensitivities of 41 microlensing events from the 2015 \emph{Spitzer} campaign, all of which received dense coverage by OGLE-IV and KMTNet. Because of the null detection of planets in this statistical sample, we provide upper limits on the planet distribution function (Equation~\ref{eqn:pdf}). In particular, we find that the normalization factor $\mathcal{A}<0.49$ at 95\% confidence level for a flat planet mass function. These constraints are consistent with the previous microlensing results by \citet{Gould:2010,Shvartzvald:2016} and \citet{Suzuki:2016}.

We develop the methodology to statistically study the Galactic distribution of planets using microlensing parallax measurements. In particular, we provide mathematical descriptions for estimating the lens mass $M_\l$ and distance parameter $\Dparm$ with the measurement of the microlensing parallax vector $\bdv{\pi_\e}$. Although such statistical estimates cannot be used as deterministic measurements of individual microlenses, they are in general fairly compact and independent of the details of the input Galactic model, because of the kinematic information contained by $\bdv{\pi_\e}$ (Figure~\ref{fig:prior}, see also \citealt{Han:1995}). In fact, the majority of events in our raw sample have uncertainties on the distance parameter, $\sigma(\Dparm)\lesssim 1~$kpc. For the purpose of determining a Galactic distribution of planets, we decide to use $\sigma(\Dparm)<1.4~$kpc as the criterion for claiming a good parallax measurement. Note that this criterion is formed based on a planet-free sample, meaning that it is not biased by the presence of any planet detection. Events that show planetary perturbations, however, do have smaller uncertainties on the lens distance parameter. This is partly because of the break down of the four-fold degeneracy, but mostly because most of them show the finite-source effect \citep{Zhu:2014}, which, when combined with the microlensing parallax measurement, yields deterministic lens distance and mass measurements \citep[e.g.,][]{Street:2016}. Therefore, while our current sample is planet-free, we suggest that the inclusion of any future planetary event into the statistical sample should be based on the $\sigma(\Dparm)$ that is estimated in the same way as a single-lens event, rather than the $\sigma(\Dparm)$ that is determined by combining information from the planetary anomaly (e.g., the finite-source effect). 

We use one of the published planetary events from the 2015 \emph{Spitzer} campaign, OGLE-2015-BLG-0966 \citep{Street:2016}, as an example to demonstrate whether a planet can be included in the sample or not. 
To remove the influence of the planet, we replace those data points that are affected by the planet with pseudo data points that are generated based on the single-lens model.
\footnote{We use the non-planetary parameters of the planetary model as parameters for this single-lens model.}
We then search for lens parameters of four degenerate solutions. The lens distance parameter $\Dparm$ is then estimated following the equations in Section~\ref{sec:bayes}. From this we then determine the median and the half-width of the 68\% confidence interval, and find $\Dparm=3.1\pm1.2~$kpc. According to the criterion $\sigma(\Dparm)<1.4~$kpc, the associated planet would be included in the statistical sample if this event had been covered by KMTNet.
\footnote{In fact, it fell in a gap between CCD chips of the camera, which would not have occurred under the 2016 KMTNet observing strategy.}

We note that we developed the criterion for ``measured $\bdv{\pi_\e}$'' before looking at OGLE-2015-BLG-0966 (or any other \emph{Spitzer} planetary event), precisely to allow us to advocate for this criterion without in any way being influenced by subconscious desire to include more planets in the sample.

Furthermore, high-magnification events such as OGLE-2015-BLG-0966 provide additional constraints on the lens mass and distance. Even if the lens is a point mass, if the lens transits the face of the source (i.e. the source crosses the caustic of a point mass, which is a single point), the event will show the finite-source effect. In the presence of the finite-source effect, the angular Einstein radius and thus the lens mass and distance are directly measured given a measurement of the parallax \citep{Yoo:2004}. Furthermore, the absence of finite-source effects provides an upper limit on the scaled source radius $\rho \lesssim u_0$, which corresponds to a lower limit on $\pi_\rel$ and an upper limit on $\Dparm$. In this way, this additional constraint reduces the uncertainty on the distance parameter $\Dparm$ and thus potentially increases the chance for high-magnification events to be included in the statistical sample. This is important because high-magnification events have much higher sensitivity to planets compared to typical events \citep{Griest:1998}. It is nevertheless unbiased in terms of planet detections, since the additional information used here does not rely on the presence of planets. In the case of OGLE-2015-BLG-0966, this additional constraint yields $\Dparm<6.9~$kpc, and thus reduces $\sigma(\Dparm)$ to 1.1~kpc.

Based on the current sample, we find that $\sim$1/3 of all planet sensitivities come from events in the bulge. Assuming the planet distribution is the same in the bulge as in the disk, this result predicts that $\sim$1/3 of all planet detections from our experiment will be in the bulge. In the future, deviations from this prediction can then be used to constrain the abundance of planets in the bulge relative to the disk.

\begin{deluxetable*}{cclllllllll}
    \tablecaption{Best-fit parameters and associated uncertainties for the 50 events in the raw sample. For each event, we only include solutions that have $\Delta\chi^2<100$ from the lowest value. We present the total baseline magnitude in OGLE-IV $I$ band, $I_{\rm base}$, the blending fraction in $I$ band, and the source $I-[3.6\ \microm]$ color, rather than the flux parameters $(F_{\rm s},F_{\rm b})$ of individual data sets. We assume no blending for events OGLE-2015-BLG-(0011, 0029, 0081, 0772, 0798, 1096, 1188, 1289, 1297, 1341, 1470), because free blending would lead to severely negative blending.
\label{tab:parameters}}
\tabletypesize{\scriptsize}
\tablehead{\colhead{OGLE \#} & \colhead{Solution} & \colhead{$\Delta\chi^2$} & \colhead{$t_0$} & \colhead{$u_0$} & \colhead{$t_\e$} & \colhead{$\pi_{\rm E,N}$} & \colhead{$\pi_{\rm E,E}$} & \colhead{$I_{\rm base}$$^a$} & \colhead{Blending} & \colhead{$I-[3.6\microm]$}}
\startdata
0011 & $(+,+)$ & 0.0 & 7125.99(12) & 0.8919(5) & 80.96(26) & 0.262(6) & 0.104(4) & 15.948 & \nodata & 2.494(26)\\
 & $(-,-)$ & 31.3 & 7125.53(10) & -0.8919(5) & 74.0(3) & -0.403(9) & 0.089(4) & 15.948 & \nodata & 2.529(24)\\
0029 & $(+,+)$ & 0.0 & 7233.03(7) & 0.7603(3) & 101.77(11) & -0.1199(17) & -0.1192(17) & 15.132 & \nodata & 2.005(12)\\
0034 & $(+,+)$ & 0.0 & 7140.37(3) & 0.282(4) & 64.2(4) & 0.055(29) & -0.057(5) & 16.559 & 0.009(18) & 3.06(4)\\
 & $(-,-)$ & 1.4 & 7140.37(3) & -0.283(3) & 63.3(5) & -0.13(5) & -0.061(4) & 16.559 & 0.001(15) & 3.06(4)\\
0081 & $(+,+)$ & 30.3 & 7135.4(4) & 0.7014(14) & 85.5(14) & -0.345(25) & -0.090(18) & 17.892 & \nodata & 2.758(14)\\
 & $(-,-)$ & 0.0 & 7135.7(5) & -0.7037(13) & 97.7(24) & 0.475(28) & -0.130(21) & 17.892 & \nodata & 2.74(3)\\
0350 & $(+,+)$ & 5.1 & 7214.86(12) & 0.445(19) & 70.7(20) & -0.049(3) & -0.068(7) & 17.295 & 0.12(5) & 2.626(21)\\
 & $(-,-)$ & 0.0 & 7214.99(12) & -0.479(19) & 66.7(18) & 0.050(4) & -0.090(8) & 17.295 & 0.01(5) & 2.617(23)\\
 & $(-,+)$ & 19.4 & 7214.50(11) & -0.448(17) & 65.0(17) & 0.567(20) & -0.205(9) & 17.295 & 0.11(4) & 2.595(23)\\
0379 & $(+,+)$ & 2.8 & 7189.62(6) & 0.448(15) & 64.4(14) & 0.003(11) & -0.098(7) & 18.087 & 0.18(3) & 2.61(3)\\
 & $(-,-)$ & 0.0 & 7189.63(6) & -0.461(15) & 62.7(14) & -0.001(13) & -0.119(7) & 18.087 & 0.13(4) & 2.59(5)\\
0388 & $(+,+)$ & 0.0 & 7161.55(7) & 0.508(21) & 40.3(11) & 0.11(3) & 0.048(29) & 17.425 & 0.25(4) & 3.67(6)\\
 & $(-,-)$ & 2.2 & 7161.55(7) & -0.518(23) & 39.8(12) & -0.15(5) & 0.018(29) & 17.425 & 0.25(4) & 3.65(6)\\
 & $(-,+)$ & 2.1 & 7161.66(9) & -0.478(18) & 46.5(11) & 1.40(7) & -0.27(5) & 17.425 & 0.32(3) & 3.63(5)\\
0461 & $(+,+)$ & 1.8 & 7161.27(19) & 0.87(5) & 41.6(17) & -0.20(5) & 0.05(4) & 17.414 & -0.01(10) & 1.76(7)\\
 & $(-,-)$ & 0.0 & 7161.31(20) & -0.86(6) & 42.9(23) & 0.31(8) & 0.05(4) & 17.414 & 0.04(10) & 1.71(9)\\

\enddata
\tablecomments{$^a$ The uncertainty of $I_{\rm base}$ is $\sim1$ mmag, primarily arising from OGLE-IV's data recording format. In addition, the calibration precision of OGLE-IV $I$ band to the standard system, $\sim10$ mmag, is not included here, on the base that it does not affect the determination of microlensing parameters.\\
$^b$ This table is available in its entirety in the machine-readable format.}
\end{deluxetable*}

\acknowledgements
Work by WZ and AG was supported by NSF grant AST-1516842. Work by AG was also supported by JPL grant 1500811.
Work by C. Han was supported by the Creative Research Initiative  Program  (2009-0081561)  of  National  Research Foundation of Korea.
The OGLE project has received funding from the National Science Centre, Poland, grant MAESTRO 2014/14/A/ST9/00121 to A.U.. 
This work is based in part on observations made with the \emph{Spitzer} Space Telescope, which is operated by the Jet Propulsion Laboratory, California Institute of Technology under a contract with NASA. 
This research has made use of the KMTNet system operated by KASI and the data were obtained at three host sites of CTIO in Chile, SAAO in South Africa, and SSO in Australia.

\appendix
\section{Source Distance Bias} \label{sec:gamma}
We parametrize the luminosity function of bulge stars in $I$ band given by \citet{Holtzman:1998} as
\begin{equation} \label{eqn:holtzman-lf}
    \log{N} = a M_I + b = \left\{ \begin{array}{ll}
            0.57M_I+0.81 & ,\ (M_I<3.5) \cr
            0.16M_I+2.24 & ,\ (M_I>3.5) \cr
        \end{array} \right.\ ,
\end{equation}
where $N$ is the number of stars per sq. arcmin per magnitude. 
In terms of the microlensing observable, $I_\s$, which is the source apparent $I$ magnitude at baseline, the number of stars per unit area per magnitude is then
\begin{equation} \label{eqn:Nstar}
    N = 10^{a(I_\s-A_I-5\log{D_\s}+5)+b}\ .
\end{equation}
Here $A_I$ is the extinction to the source. For given $I_\s$ and $D_\s$, $a$ and $b$ can be determined by comparing the derived $M_I$ with the magnitude threshold in Equation~(\ref{eqn:holtzman-lf}). 

In principle, one should use the full expression of $N$ given in Equation~(\ref{eqn:Nstar}) as the third factor in the weight of $D_\s$ (Equations~\ref{eqn:ddist-prior} and \ref{eqn:bulge-frac}). This is because the values of $a$ and $b$ may change as $D_\s$ varies. However, with the extinction map given in \citet{Nataf:2013}, we find that nearly all sources in our sample have $M_I$ considerably below 3.5 for typical $D_\s\sim8.3~$kpc, so we use the simplified weight $D_\s^{-\gamma}$ and choose $\gamma=5\times0.57=2.85$. 

\begin{figure}
    \epsscale{1.2}
    \plotone{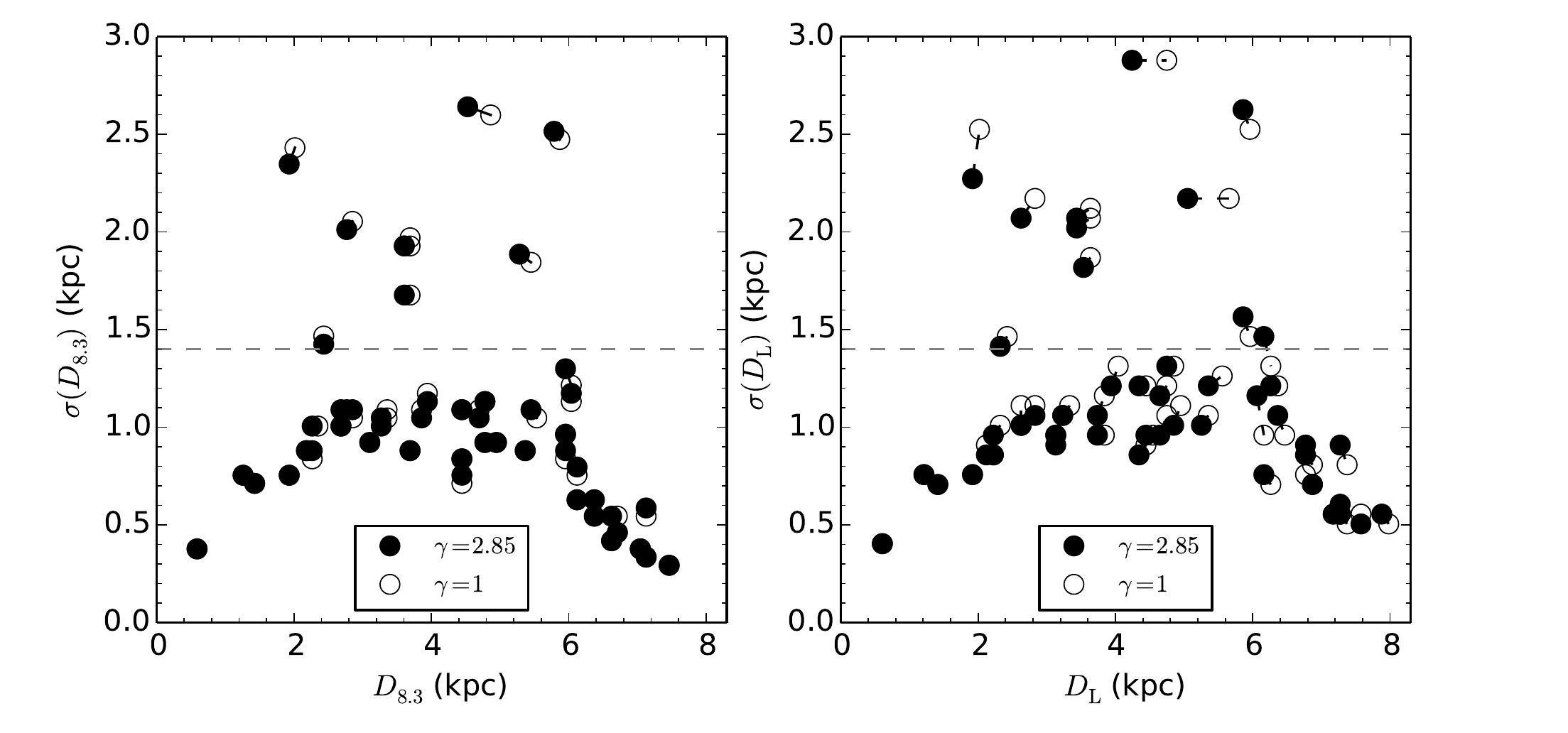}
    \caption{The impact of $\gamma$ in median and dispersion of the lens distance parameter $\Dparm$ (left panel) and $D_\l$ for the 50 events in the raw sample. The gray dashed line is the threshold for claiming a ``good'' parallax measurement. See Figure~\ref{fig:measure-pie} for more details.
    \label{fig:gamma-effect}}
\end{figure}

We note that the resulting lens distributions are insensitive to the choice of $\gamma$. For example, the variation in the $\Dparm$ distribution derived by Equation~(\ref{eqn:ddist-prior}) is limited to within 5\% if $\gamma=1$ is used, and the shift in the median of $\Dparm$ is $\lesssim0.1$~kpc (see the left panel of Figure~\ref{fig:gamma-effect}). This is a consequence of three factors. First, the number density term, $n_\s$, dominates over $D_\s^{2-\gamma}$, so that the mean source distance, $\langle D_\s \rangle \equiv (\int n_\s D_\s^{3-\gamma} \dif D_\s)/(\int n_\s D_\s^{2-\gamma} \dif D_\s)$, only differs by $0.16~$kpc when $\gamma$ changes from 2.85 to 1. Second, we derive the lens position in terms of $\Dparm$ rather than the actual lens distance $D_\l$, and $\Dparm$ is less dependent on $D_\s$ than $D_\l$ is. To further demonstrate this point, we also derive the distribution of $D_\l$, which involves
\begin{equation} \label{eqn:bayesian-dl}
    \frac{\dif^4\Gamma}{\dif D_\l \dif t_\e^\prime \dif^2\bdv{\tilde{v}_\hel}} = 4n_\l D_\l^2 f_{\tilde{v}}(\bdv{\tilde{v}_\hel}) \frac{\dif \xi(M_\l)}{\dif \log M_\l} \mu_\rel^2\ .
\end{equation}
As shown in Figure~\ref{fig:gamma-effect}, different choices of $\gamma$ can lead $D_\l$ to differ by $\sim$0.1~kpc for $D_\l \gtrsim2~$kpc, but the difference in $\Dparm$ is in general much smaller, and only reaches $\sim$0.1~kpc when $\Dparm\sim4$~kpc.

The third reason is that, although the solution based on $\bdv{\pi_\e}$ measurement is fairly compact, the dispersion is still considerably large compared to $\sim$0.1~kpc. As an extreme example, if the lens distribution for a fixed source distance, $\mathcal{P}(\Dparm|D_\s)$ (or $\mathcal{P}(D_\l|D_\s)$), is perfectly uniform, $\gamma$ will have no impact on the result at all, because terms containing $\gamma$ in Equation~(\ref{eqn:ddist-prior}) cancel out.

\begin{figure}
    \epsscale{0.8}
    \plotone{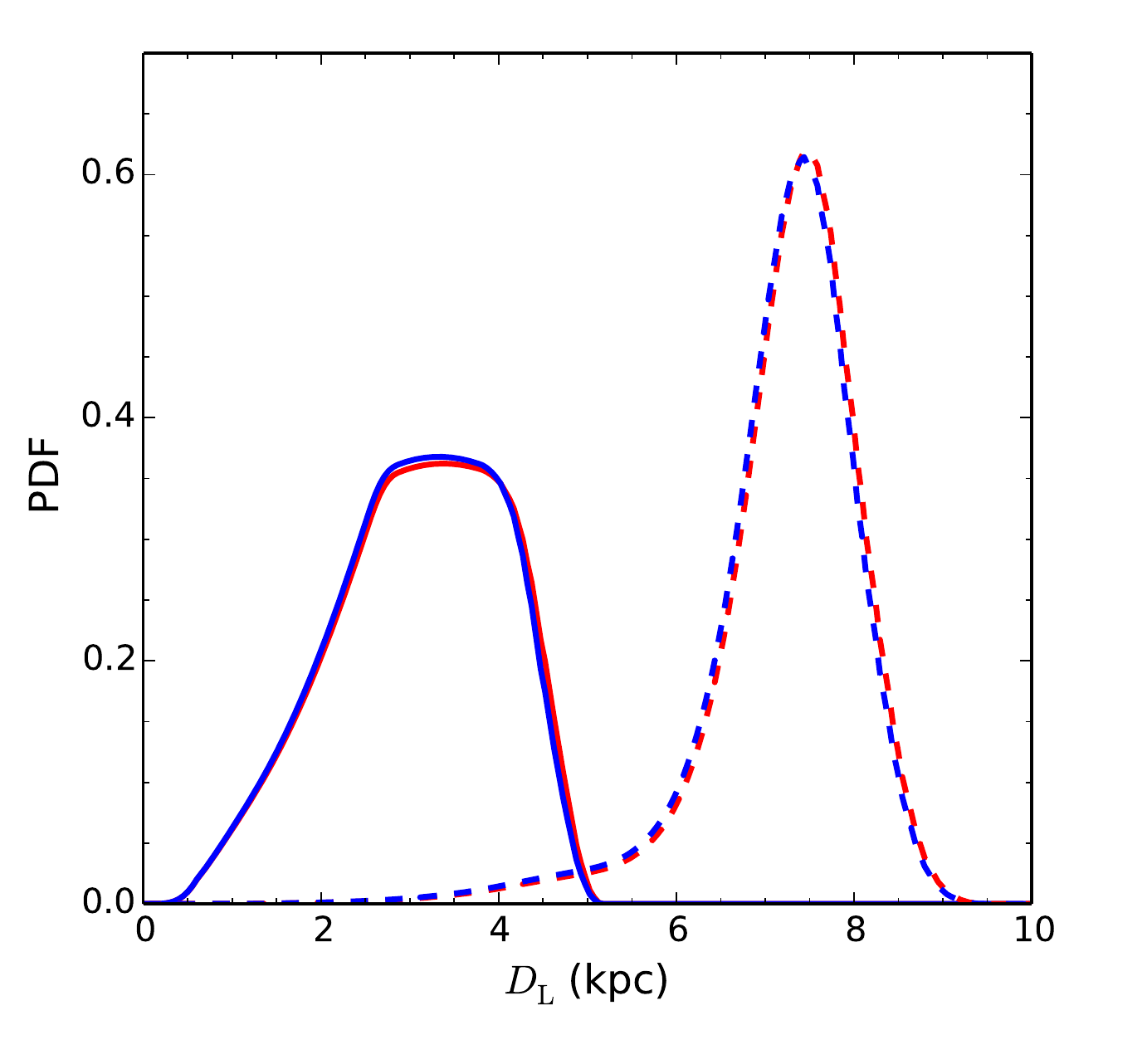}
    \caption{Lens distance distributions of OGLE-2005-BLG-169 (solid curves) and MOA-2011-BLG-293 (dashed curves) based on $\theta_\e$ measurements. Results with $\gamma=1$ are shown in red, and those with $\gamma=2.85$ are shown in blue. The difference in the median of $D_\l$ is less than $0.05$~kpc. In the derivation, a Kroupa-like mass function (Equation~\ref{eqn:kroupa}) has been assumed. We also assume perfect knowledge of $\theta_\e$ and $t_\e$, meaning that $P(\theta_\e|\data)$ and $P(t_\e|\data)$ in Equation~\ref{eqn:bayes-thetae} are both Dirac-$\delta$ functions.
    \label{fig:thetae}}
\end{figure}

This argument also implies that $\gamma$ has even smaller effect for the Bayesian estimates of lens distances based on $\theta_\e$ measurements, which in general have broader distributions. To prove this point, we first provide the corresponding $\mathcal{P}(D_\l|D_\s)$ in the case that $\theta_\e$ rather than $\bdv{\pi_\e}$ is measured,
\footnote{Note that now the correction from $t_\e$ to $t_\e^\prime$ is no longer achievable. We therefore assume $t_\e^\prime=t_\e$, which is in general a reasonable assumption (since $v_\oplus\ll \tilde{v}_\hel$).}
\begin{equation} \label{eqn:bayes-thetae}
    \mathcal{P}(D_\l|D_\s) = \int \frac{\dif^3\Gamma}{\dif D_\l \dif\theta_\e \dif t_\e} P(\theta_\e|\data)P(t_\e|\data) \dif\theta_\e \dif t_\e;\quad
    \frac{\dif^3\Gamma}{\dif D_\l \dif\theta_\e \dif t_\e} = 4n_\l D_\l^2 f_\mu(\mu_\rel) \frac{\dif\xi(M_\l)}{\dif\log M_\l} \frac{\mu_\rel^2}{t_\e}\ .
\end{equation}
Here $f_\mu(\mu_\rel)$ is the probability distribution of $\mu_\rel$, and is given by
\begin{equation}
    f_\mu(\mu_\rel) = \frac{\mu_\rel}{2\pi\sigma_l\sigma_b} \int_0^{2\pi} \exp\left[ -\frac{(\mu_\rel\cos\theta-\bar{\mu}_\rel^l)^2}{2\sigma_l^2}-\frac{(\mu_\rel\sin\theta-\bar{\mu}_\rel^b)^2}{2\sigma_b^2} \right] \dif\theta\ ,
\end{equation}
where $\bar{\mu}_\rel^l$ and $\sigma_l$ are the mean and dispersion of $\bdv{\mu_\rel}$ along the $l$ direction, and $\bar{\mu}_\rel^b$ and $\sigma_b$ are the counterparts along the $b$ direction, respectively. With these formulae, we then derive the lens distances of two published events, OGLE-2005-BLG-169 (a typical disk event, \citealt{Gould:2006,Batista:2015,Bennett:2015}) and MOA-2011-BLG-293 (a typical bulge event, \citealt{Yee:2012,Batista:2014}), and show the resulting distributions in Figure~\ref{fig:thetae}. As expected, the differences in $D_\l$ arising from different values of $\gamma$ are smaller compared to cases with $\bdv{\pi_\e}$ measurements.

\section{Integral of Gaussian Product} \label{sec:gaussian}

The estimates of lens mass $M_\l$ and distance parameter $\Dparm$ (Equation~\ref{eqn:ddist-prior}) involve the integral of the product of two multi-dimension Gaussian probability functions. Although this result has probably been well known for centuries, we provide an explicit representation below simply for completeness.

The integral of the product of two multi-dimension Gaussian distributions can be written as
\begin{equation} \label{eqn:integral}
    P(\bdv{x}) = \int \mathcal{N}(\bdv{x}|\bdv{\mu_1},\mathbb{C}_1) \mathcal{N}(\bdv{x}|\bdv{\mu_2},\mathbb{C}_2) \dif^n \bdv{x}\ .
\end{equation}
Here $\mathcal{N}(\bdv{x}|\bdv{\mu},\mathbb{C})$ is the notation for a multi-dimension Gaussian probability function with mean $\bdv{\mu}$ and covariance matrix $\mathbb{C}$,
\begin{equation}
    \mathcal{N}(\bdv{x}|\bdv{\mu},\mathbb{C}) \equiv \frac{\exp \left[-\frac{1}{2}\left(\bdv{x}-\bdv{\mu}\right)^{\rm T} \mathbb{C}^{-1} \left(\bdv{x}-\bdv{\mu}\right) \right]}{\sqrt{|2\pi \mathbb{C}|}}
\end{equation}
The integral given by Equation~(\ref{eqn:integral}) can be computed analytically by ``completing the squares'',
\begin{equation}
    P(\bdv{x}) = \sqrt{\frac{|2\pi\hat{\mathbb{C}|}}{|2\pi\mathbb{C}_1| |2\pi\mathbb{C}_2|}} \exp\left[ -\frac{1}{2} \left( \bdv{\mu_1}^{\rm T} \mathbb{C}_1^{-1}\bdv{\mu_1} + \bdv{\mu_2}^{\rm T} \mathbb{C}_2^{-1}\bdv{\mu_2} -\bdv{\hat{\mu}}^{\rm T} \hat{\mathbb{C}}^{-1} \hat{\bdv{\mu}}\right) \right]\ ,
\end{equation}
where
\begin{equation}
    \hat{\mathbb{C}}^{-1} \equiv \mathbb{C}_1^{-1} + \mathbb{C}_2^{-1};\quad \bdv{\hat{\mu}} \equiv \hat{\mathbb{C}} \left( \mathbb{C}_1^{-1}\bdv{\mu_1} + \mathbb{C}_2^{-1}\bdv{\mu_2}\right)\ .
\end{equation}

\end{CJK*}
\end{document}